\documentclass[preprint]{aastex63}

\revised{\today}
\submitjournal{AJ}

\shorttitle{X-Ray Spectral Parameter Survey}
\shortauthors{Kinch et al.}

\begin{document}

\title{Spin and Accretion Rate Dependence of Black Hole X-Ray Spectra}

\correspondingauthor{Brooks E. Kinch}
\email{kinch@lanl.gov}

\author[0000-0002-8676-425X]{Brooks E. Kinch}
\affiliation{CCS-2: Computational Physics and Methods \\
Los Alamos National Laboratory \\
NM 87545, USA}

\author[0000-0002-2942-8399]{Jeremy D. Schnittman}
\affiliation{Gravitational Astrophysics Laboratory, Code 663 \\
NASA Goddard Space Flight Center \\
Greenbelt, MD 20771, USA}

\author[0000-0003-3547-8306]{Scott C. Noble}
\affiliation{Gravitational Astrophysics Laboratory, Code 663 \\
NASA Goddard Space Flight Center \\
Greenbelt, MD 20771, USA}

\author[0000-0002-5779-6906]{Timothy R. Kallman}
\affiliation{X-Ray Astrophysics Laboratory, Code 662 \\
NASA Goddard Space Flight Center \\
Greenbelt, MD 20771, USA}

\author[0000-0002-2942-8399]{Julian H. Krolik}
\affiliation{Department of Physics and Astronomy \\
Johns Hopkins University \\
Baltimore, MD 21218, USA}

\begin{abstract}

We present a survey of how the spectral features of black hole X-ray binary systems depend on spin, accretion rate, viewing angle, and Fe abundance when predicted on the basis of first principles physical calculations. The power law component hardens with increasing spin. The thermal component strengthens with increasing accretion rate. The Compton bump is enhanced by higher accretion rate and lower spin. The Fe K$\alpha$ equivalent width grows sub-linearly with Fe abundance. Strikingly, the K$\alpha$ profile is more sensitive to accretion rate than to spin because its radial surface brightness profile is relatively flat, and higher accretion rate extends the production region to smaller radii. The overall radiative efficiency is at least 30–-100\% greater than as predicted by the Novikov-Thorne model.

\end{abstract}


\ \\
\ \\

\section{Introduction}

The X-ray photons which escape from the innermost regions of a black hole system convey information about the mass and spin of the black hole itself as well as the properties of the accretion flow in its immediate environment. X-ray spectra observed from stellar-mass black holes in X-ray binary systems typically show the following common features: a thermal peak superimposed on a power law which extends to high energy and eventually rolls over to an exponential decay, an extremely broad Fe K$\alpha$ fluorescent emission line centered at 6--7 keV, and a wide ``bump'' superimposed on the power law above the K$\alpha$ feature. In recent years, much has been learned from General Relativistic magnetohydrodynamic (GRMHD) simulations about the dynamics of the accretion flow: how it is regulated, in part, by the black hole’s mass and spin, and how, at a qualitative level, it could create these spectral features.

Despite these advances, the standard picture of the accretion flow geometry often used for data analysis is little more than a cartoon: a razor thin but optically thick disk [with inner cutoff at exactly the innermost stable circular orbit (ISCO) radius] emits blackbody radiation according to a prescribed radial temperature profile [usually that of \citet{sha73a}], while it is illuminated by hard X-rays generated in a stationary ``lamppost'' located at a single point on the rotation axis some height above the black hole. In reality, it is virtually impossible for plasma to remain stationary at such a location, and both physical and observational arguments point strongly to an extended \citep{wil16a, cha19a, zog20a, zog21a, zdz21a} and dynamic \citep{kar19a} corona. Moreover, these features are by no means independent: the hard X-rays are created by Compton upscattering of thermal disk photons in the corona, and are ``reprocessed'' if they strike the disk to create both the K$\alpha$ line and the Compton bump. The profile of the K$\alpha$ line is therefore determined by how the hard X-ray flux varies over the surface of the disk in concert with how the effectiveness of reprocessing varies over that same surface, as well as Compton scattering in the corona as they escape.


All this physics is well-understood, but the many interactions make a complex problem. In order to calculate black hole accretion self-consistently from first principles, one must contend with the relevant physics governing the processes just described: General Relativity, magnetohydrodynamics, atomic physics, relativistic Compton scattering, and radiation transport. Our method for doing so in as fundamental a way as possible is documented in \citet{kin16a, kin19a, kin20a}, and is reviewed briefly in the next section. In short, we post-process snapshots taken from state of the art three-dimensional GRMHD simulations using codes which self-consistently solve the equilibrium radiation transport and thermal plus photoionization balance problems, accounting for all relativistic effects.

In this paper, we use this established method to investigate how the features of the synthetic spectra predicted on the basis of this first principles approach depend on the properties of the modeled black hole system: the mass and spin of the black hole, the rate at which the inflowing material accretes onto the black hole, the elemental abundance of Fe, and the angle at which the disk is inclined relative to the observer's line of sight.

By beginning with self-consistent physics, this method provides a number of advantages with respect to the usual methods of spectral analysis, which involve fitting to data the model spectra generated by codes like \textsc{reflionx} \citep{ros05a}, \textsc{xillver} \citep{gar10a, gar11a, gar13a}, and \textsc{relxill} \citep{gar14a}. While the models generated by these codes do produce good fits to real spectral data, they rely on parameterized descriptions of the accretion flow geometry, illumination pattern, temperature structure, etc.---not surprisingly, they therefore require many more parameters in order to specify a spectrum. By post-processing simulation data, we not only replace an ad hoc description of the black hole environment with a physically-motivated one, we also directly connect our smaller set of parameters to the properties and physics of the real systems we seek to understand.

\section{Method}

Our starting point is the extremely well-resolved ThinHR series of simulations \citep{nob10a} performed by the 3D general relativistic MHD code \textsc{harm3d} \citep{nob09a}. These simulations studied the MHD evolution of an accretion disk in a Kerr spacetime. The matter was assumed to have an ideal gas equation of state with adiabatic index $c_P/c_V = 5/3$ [see \citet{sch13a} for a discussion concerning the choice of equation of state], cooling according to a target-temperature cooling function in the disk body \citep{nob11a} and an inverse Compton (IC) cooling function in the corona \citep{kin20a}.

The disk body radiates a roughly thermal spectrum into the hot corona, which is energized by magnetic dissipation. This thermal spectrum is Comptonized in the corona. Some of the upscattered photons strike the disk, where their energy is reprocessed, altering its quasi-thermal spectrum.

\begin{figure}
\epsscale{0.6}
\plotone{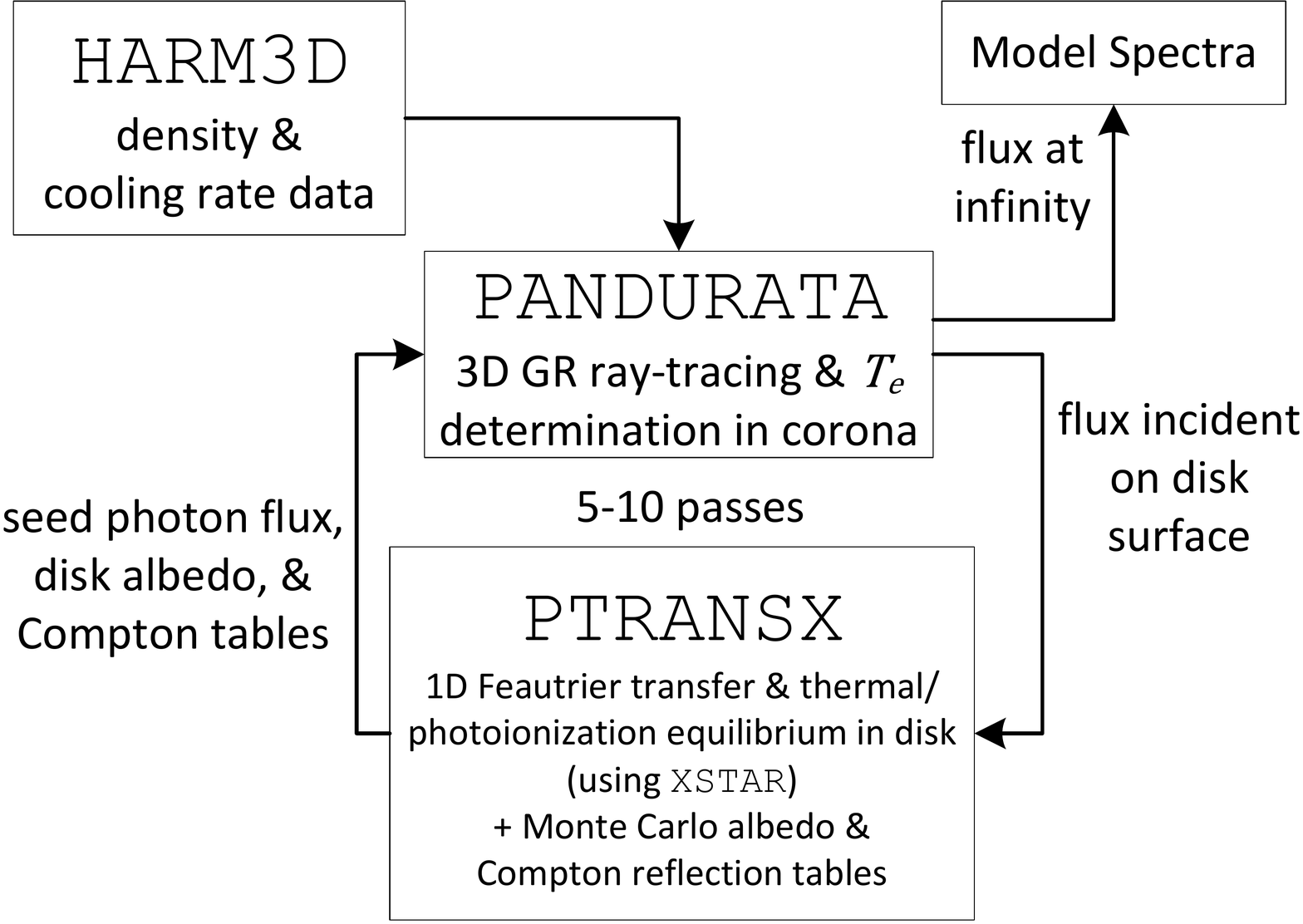}
\caption{A schematic representation of our \textsc{pandurata}+\textsc{ptransx} pipeline, which generates synthetic model spectra out of \textsc{harm3d} simulation data.
\label{fig:flowchart}}
\end{figure}

To predict the shape of the spectrum emerging from this complex feedback system, we post-processed snapshot output data using the codes \textsc{pandurata} and \textsc{ptransx} as described in \citet{kin19a}: the geodesic-integrating, ray-tracing code \textsc{pandurata} \citep{sch13b} solves the time-independent radiation transport problem in the optically thin, hot corona; for the denser, cooler disk, the plane-parallel radiative transfer solver \textsc{ptransx} \citep{kin16a}, together with \textsc{xstar} \citep{kal01a}, finds the thermal- plus photoionization-balancing seed photon spectrum and disk albedo. Applied iteratively, these codes build up a globally self-consistent solution for the radiation field, temperature structure, and ionization balance of the computational volume, accounting for all relativistic effects, and including the spectrum as would be seen by a distant observer. A schematic summary of the simulation-to-spectrum pipeline is shown in Figure \ref{fig:flowchart}.

This technique is, at present, the most realistic, first principles-based means by which to generate theoretical predictions for the X-ray spectra of black hole systems. Furthermore it is the \emph{only} method which does so starting from a description of the accretion flow environment as it develops within a magnetorotational instability (MRI)-resolving \citep{bal91a, haw91a} simulation---widely agreed to be the actual mechanism for angular momentum transport in real disks---while also reproducing all the hallmark spectral features observed in real moderately-accreting black holes. While \citet{nar16a} also generated synthetic distant observer spectra starting from simulation output, they did so without the machinery required to incorporate disk reprocessing effects (e.g., K$\alpha$ lines), and their base simulations had too little coronal luminosity to produce substantial power law components in their final spectra.

Each synthetic spectrum our method generates is characterized entirely by only five parameters: the central black hole mass ($M$), the black hole's dimensionless spin parameter ($a$; sometimes written $a^*$ or $a/M$) the mass accretion rate expressed in Eddington units ($\dot{m} = \dot{M}/\dot{M}_\mathrm{Edd}$), the Fe abundance relative to its solar value ($[\mathrm{Fe}]/[\mathrm{Fe}]_\odot$; in principle the entire elemental abundance list is a free parameter, but Fe contributes the most easily observable line emission in X-ray binary spectra), and the viewer inclination angle ($i$, where $i = 0^\circ$ is a face-on view of the disk, while $i = 90^\circ$ is an edge-on view). These parameters directly quantify independent physical or geometric properties of real black hole systems, and therefore cannot be reduced to an even smaller set. The process by which scale free simulation results are translated into physical (cgs) units for post-processing is thoroughly detailed in \citet{kin19a, kin20a}: for the purposes of this paper, it suffices to explain that the ThinHR simulations are run choosing only a value for the dimensionless spin $a$ until they achieve approximate inflow equilibrium. At that point, the accretion rate must be chosen in order to employ the IC cooling function, and the simulations are run for an additional $1000M$ to achieve a new quasi-steady state. A choice for $M$ at this stage allows the full translation of these last $1000M$ of simulation data into cgs units for post-processing, which also requires specification of the Fe abundance. Finally, because $\textsc{pandurata}$ is an emitter-to-observer ray-tracing code, spectra for all viewing angles $i$ are computed simultaneously.

An important technical caveat we briefly discuss here [again see \citet{kin20a} for details] are \textsc{harm3d}'s ``zero cooling'' conditions. Each time step, \textsc{harm3d} must recover a series of familiar  ``primitive'' variables (density, internal energy, pressure, etc.) from a more natural (in GR) set of ``conserved'' variables---those quantities which are actually evolved by \textsc{harm3d}. The conserved-to-primitive recovery procedure is performed via a nonlinear solver, and is subject to failure in extremely magnetically-dominated cells. In these regions, a separate entropy-conserving integration is performed. The thresholds beyond which we choose to invoke the entropy equation are (in dimensionless code units) $B^2/u > 10^4$ (i.e, below a minimum plasma-$\beta =(c_P/c_V-1)8\pi 10^{-4}$, where $c_P/c_V$ is the adiabatic index) or $B^2/\rho > 1$, where $B^2$ is the magnetic energy density, $u$ is the internal energy density, and $\rho$ is the (code units) density. In practice, these conditions occur only near the evacuated polar ``jet'' region of the simulations. In addition, while these simulations are spatially well-resolved in the disk and corona, the rapid changes in density and temperature as a function of polar angle approaching the $z$-axis are difficult to resolve with any standard grid [see \citet{kin20a} for details]. The net result of these limitations are patches of uncooled and often unphysically hot material near the $z$-axis. However, as we will demonstrate below, the unreliable thermodynamic treatment of the jet gas \emph{does not} affect the synthetic X-ray spectrum we generate from a given set of simulation data. These choices are made for the stability of the GRMHD integrator, and these thresholds are not part of our parameter set because they do not affect the spectral region of interest. A future study, focusing on emission from the jet region, would require a more physically robust treatment thereof: pair production must be included to correctly treat the thermodynamics of the extremely hot, diffuse plasma [see \citet{won21a} for an implementation in a \textsc{harm}-like code]; additionally, a scheme like \textsc{patchwork} \citep{shi18a} might be used to better resolve the jet/corona boundary.

\section{Results}

\subsection{Simulation Output}

\begin{figure}
\epsscale{0.9}
\plotone{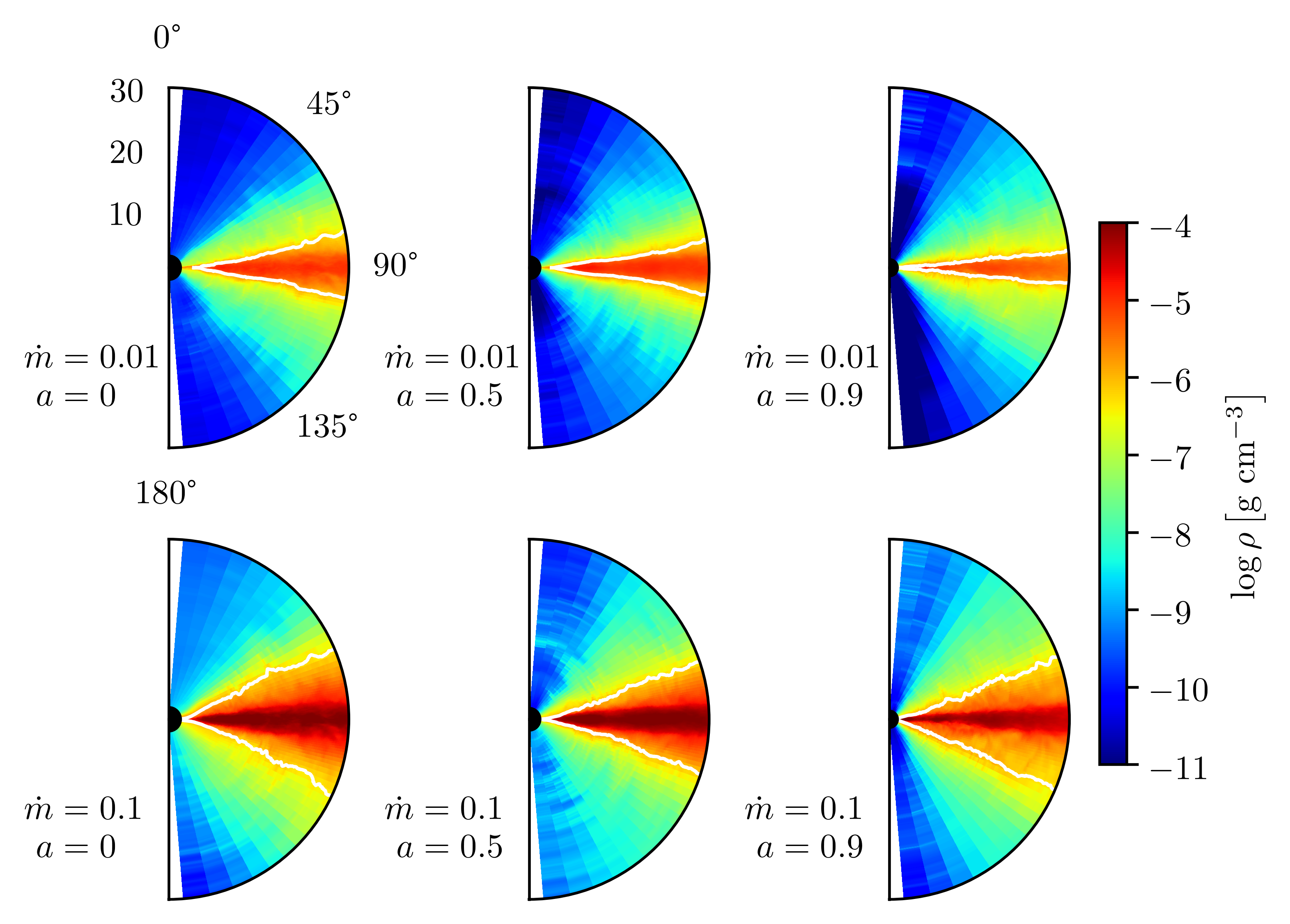}
\caption{The $\phi$-averaged density structure of each simulation, using a snapshot $1000M$ after switching on the IC cooling function. The white lines indicate the upper and lower photosphere surfaces. The simulations are run with the accretion rate and spin indicated, all with a central black hole mass of $10M_\odot$. The radial extent of these figures is $30M$, and each are plotted using the same color scale.
\label{fig:harm_rho_grid}}
\end{figure}

\begin{figure}
\epsscale{0.9}
\plotone{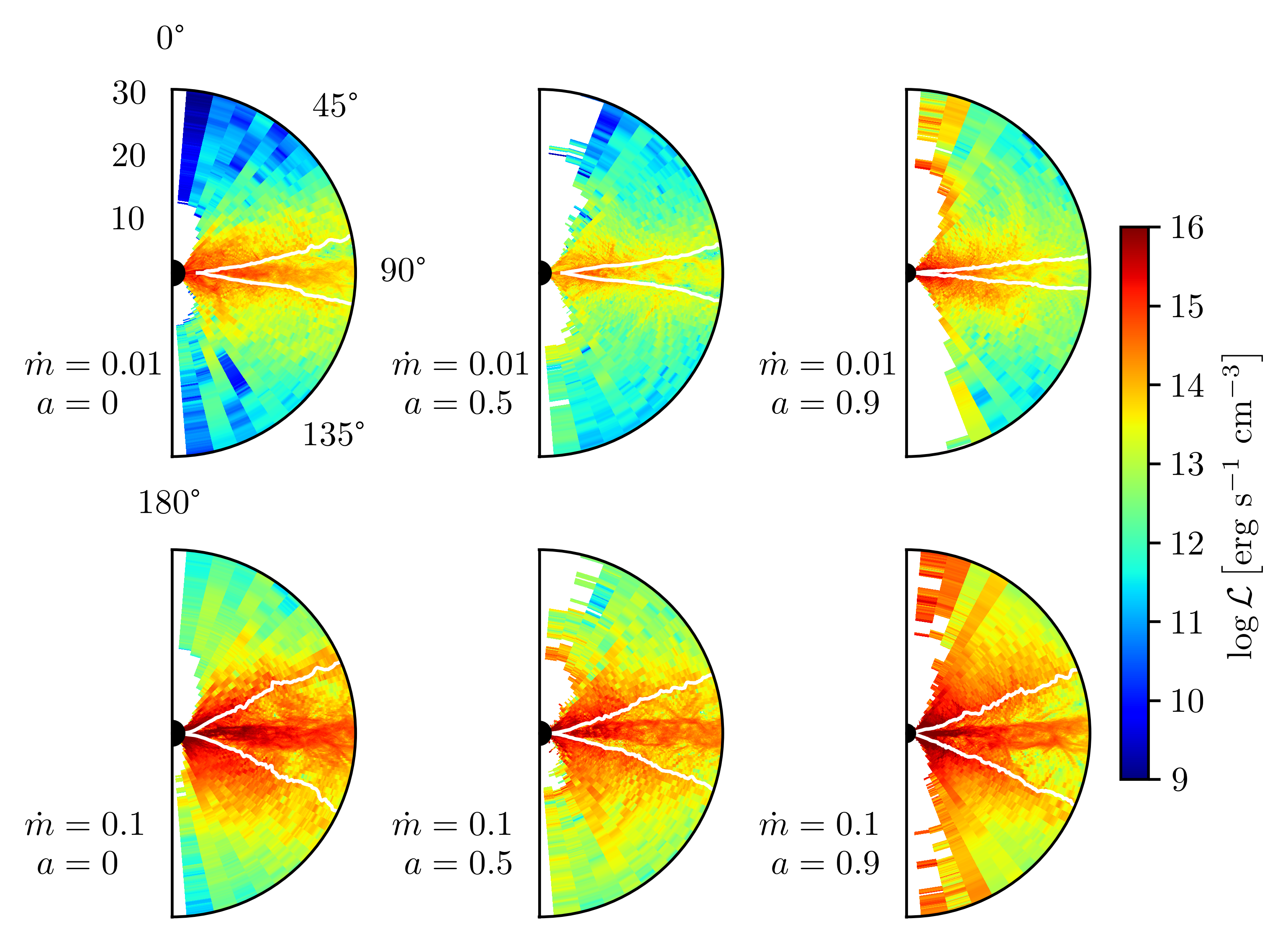}
\caption{The $\phi$-averaged cooling rate for each simulation at $t = +1000M$. The cooling rate used is the target-temperature rate in the disk (between the photosphere surfaces) and the IC cooling rate in the corona (outside the photosphere surfaces). The white regions indicate zero cooling \emph{in all $\phi$ cells} for this snapshot.
\label{fig:harm_cool_grid}}
\end{figure}

\begin{figure}
\epsscale{0.9}
\plotone{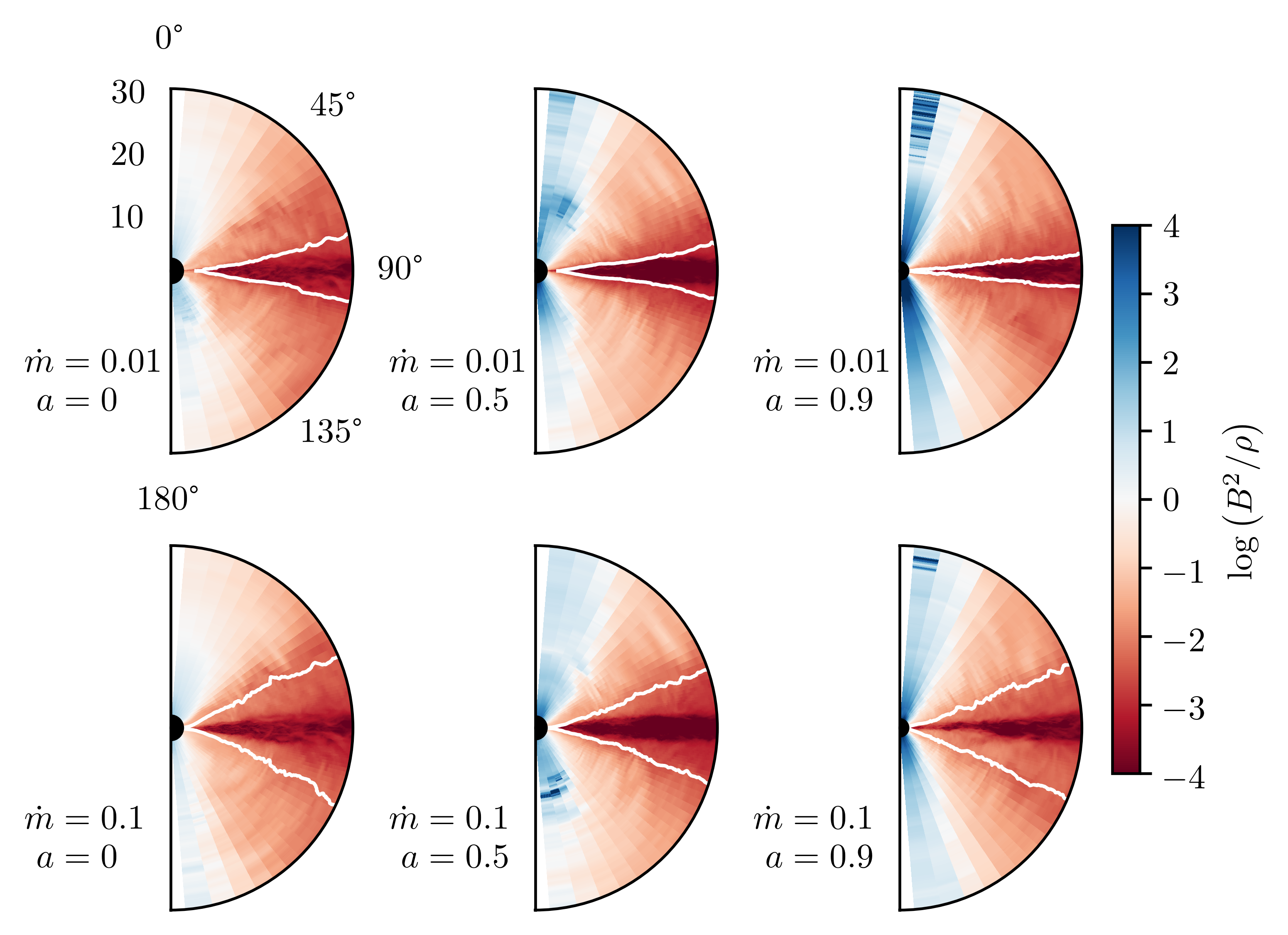}
\caption{The $\phi$-averaged magnetization (magnetic energy density divided by mass-energy density) for each simulation at $t = +1000M$. On this scale, blue indicates $\langle B^2/\rho \rangle > 1$, where we expect many cells to have zeroed-out cooling rates.
\label{fig:harm_bsq_o_rho_grid}}
\end{figure}

\begin{figure}
\epsscale{0.9}
\plotone{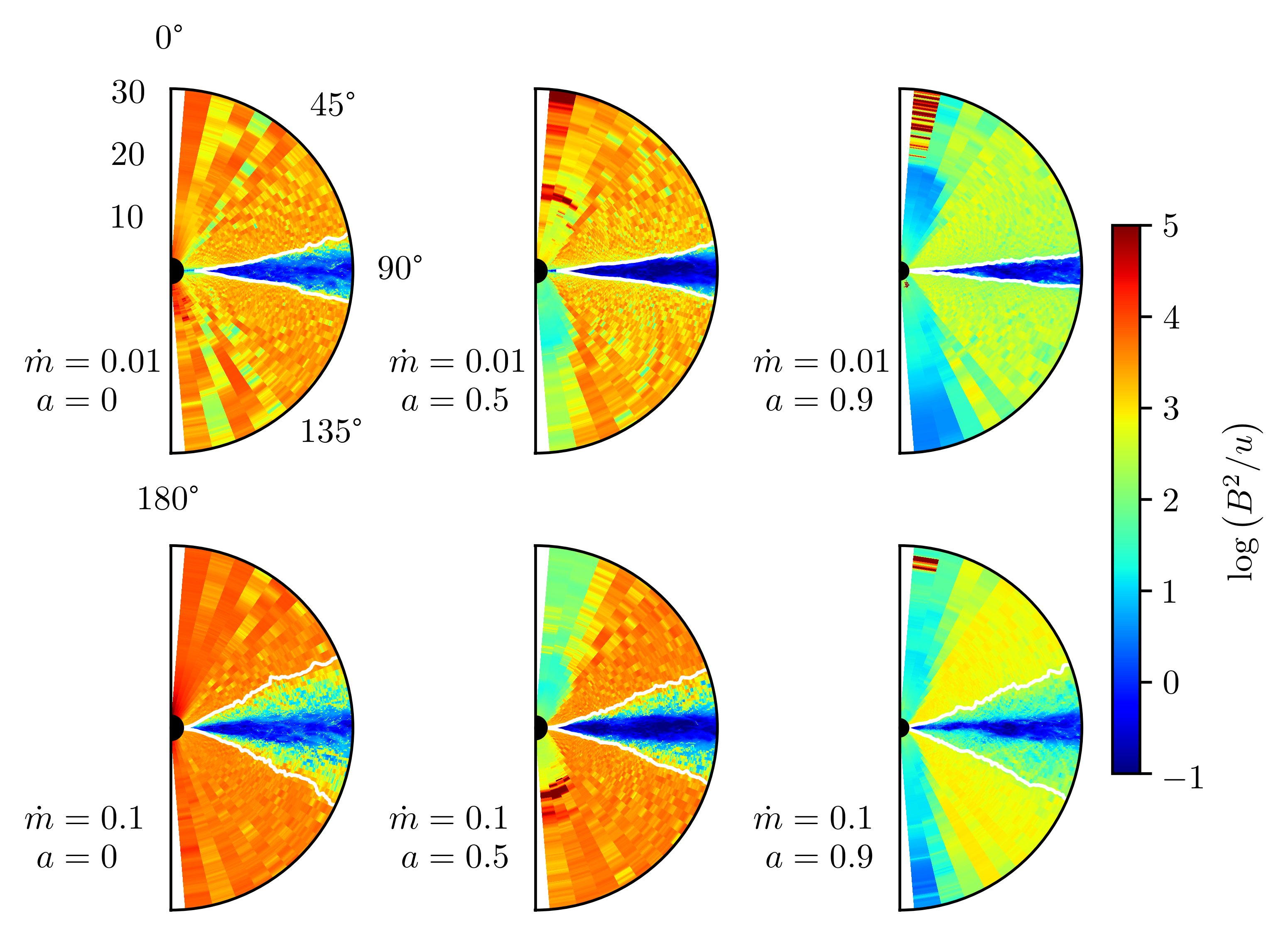}
\caption{The $\phi$-averaged ratio of magnetic energy density to internal energy density (equal to $(c_P/c_V-1)8\pi$ divided by the plasma-$\beta$) for each simulation at $t = +1000M$.
\label{fig:harm_bsq_o_u_grid}}
\end{figure}

\begin{figure}
\epsscale{0.9}
\plotone{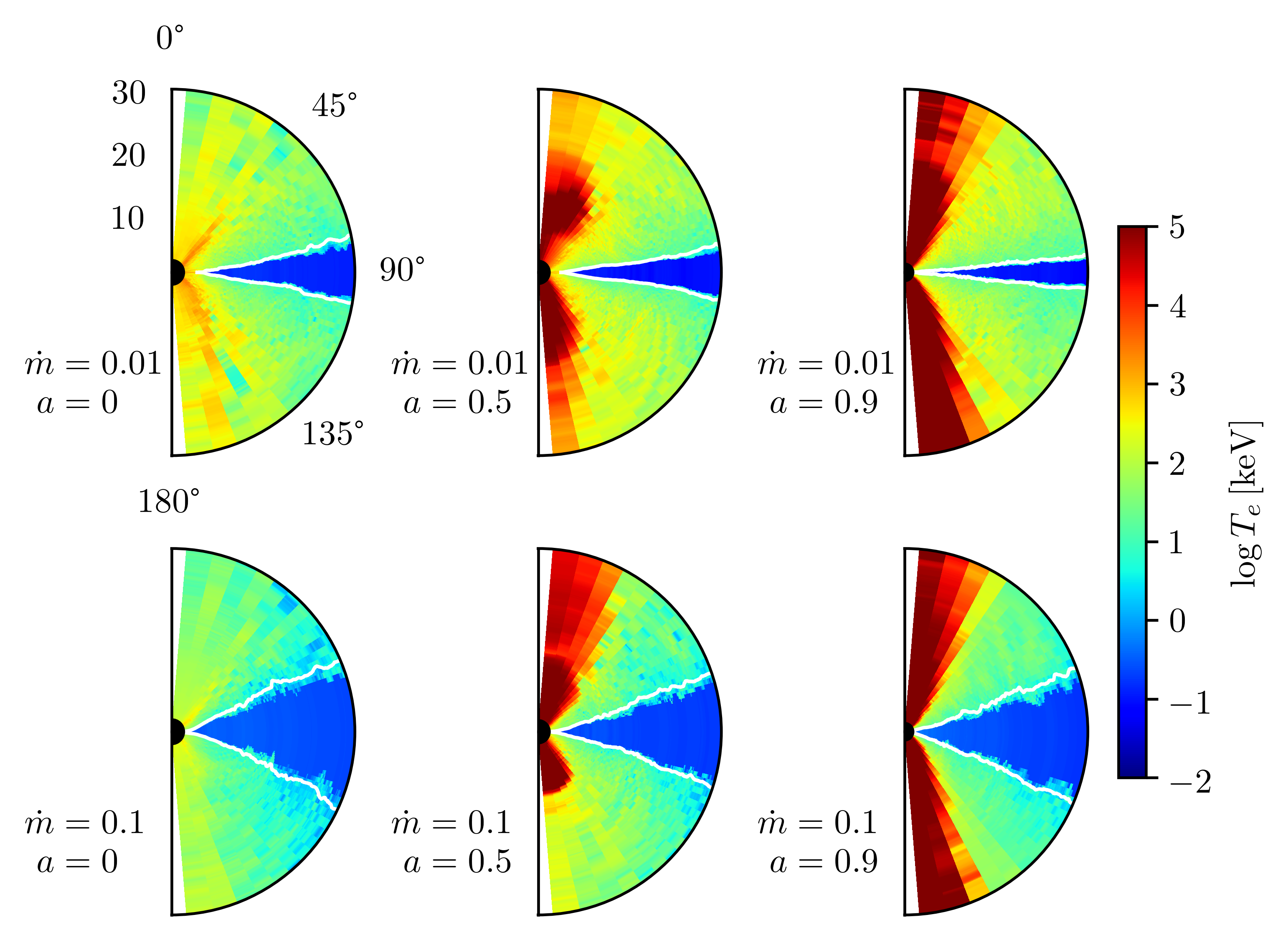}
\caption{The $\phi$-averaged electron temperature for each simulation at $t = +1000M$. For cells within the disk body, the effective temperature at the corresponding photosphere is used for $T_e$, adjusted according to our detailed photoionization calculation just inside the photosphere.
\label{fig:harm_Te_grid}}
\end{figure}

In Figures \ref{fig:harm_rho_grid}--\ref{fig:harm_Te_grid}, we show various azimuthally-averaged simulation quantities $1000M$ after switching on the IC coronal cooling function; also shown is the $\phi$-averaged location of the upper and lower photosphere surfaces. Note that as spin increases, the photosphere surfaces move closer to the midplane. This is due, in part, to how the density is scaled from dimensionless ``code'' units to physical (cgs) units: $\rho \propto \dot{m}/\eta_\mathrm{NT}$, where $\dot{m}$ is the accretion rate in Eddington units and $\eta_\mathrm{NT}$ is the \citet{nov73a} radiative efficiency ($L = \eta \dot{M} c^2$) calculated using analytic accretion disk theory, assuming a zero torque boundary condition at the ISCO. See Table 1 for the values of $\eta_\mathrm{NT}$ for the spins considered here.

As the spin increases, the density in the jet cone decreases (Figure \ref{fig:harm_rho_grid}), leading to an increasing number of coronal cells subject to evolution via the entropy equation and therefore not cooled (Figure \ref{fig:harm_cool_grid}). This is due to an increase in the magnetization $B^2/\rho$ as shown in Figure \ref{fig:harm_bsq_o_rho_grid}; the plasma-$\beta$ in fact increases with increasing spin, shown (as its reciprocal quantity $B^2/u$) in Figure \ref{fig:harm_bsq_o_u_grid}. The net result is that the jet cone becomes more unphysically hot as the spin increases, as shown in Figure \ref{fig:harm_Te_grid}.

\begin{figure}
\plottwo{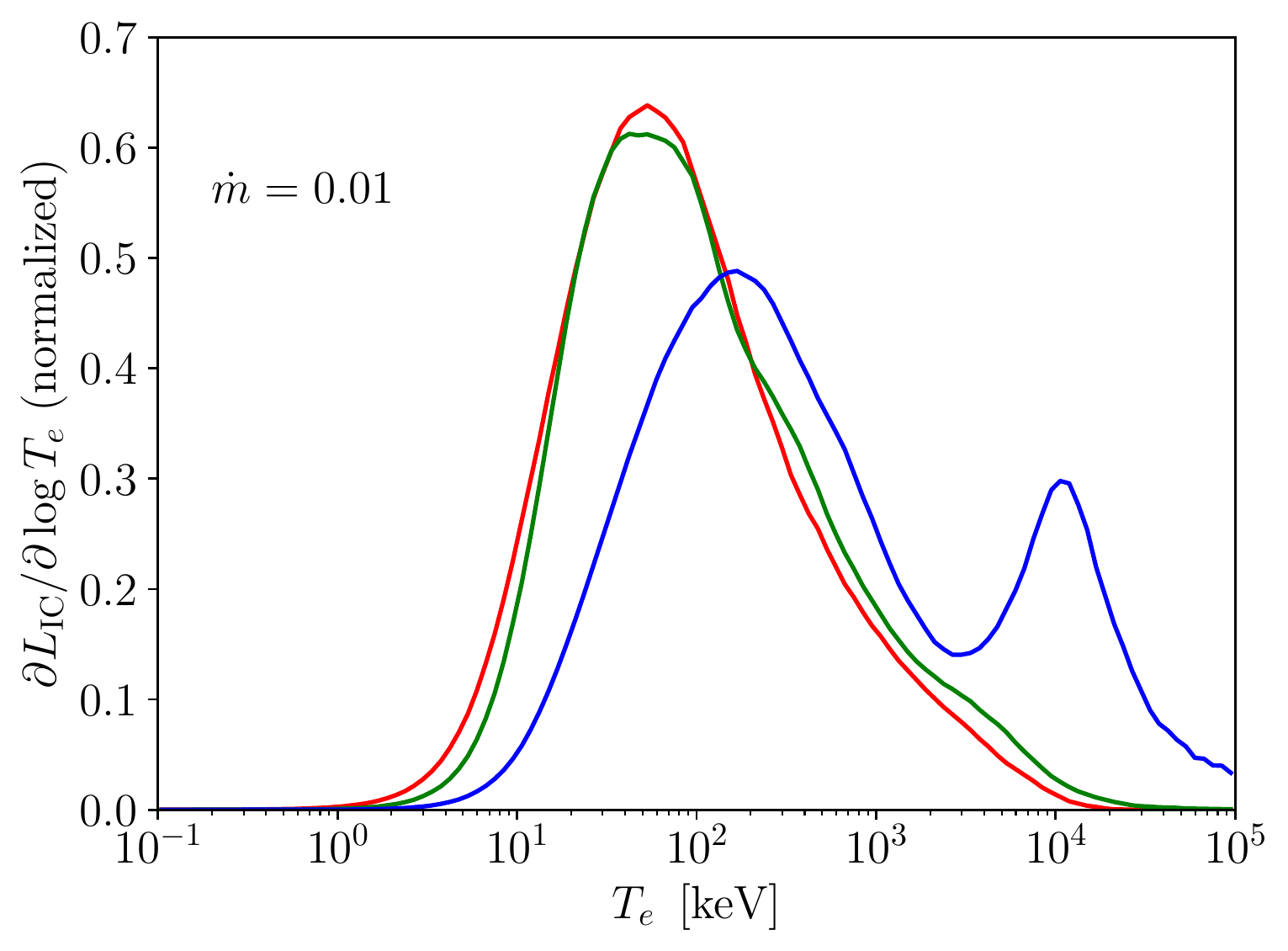}{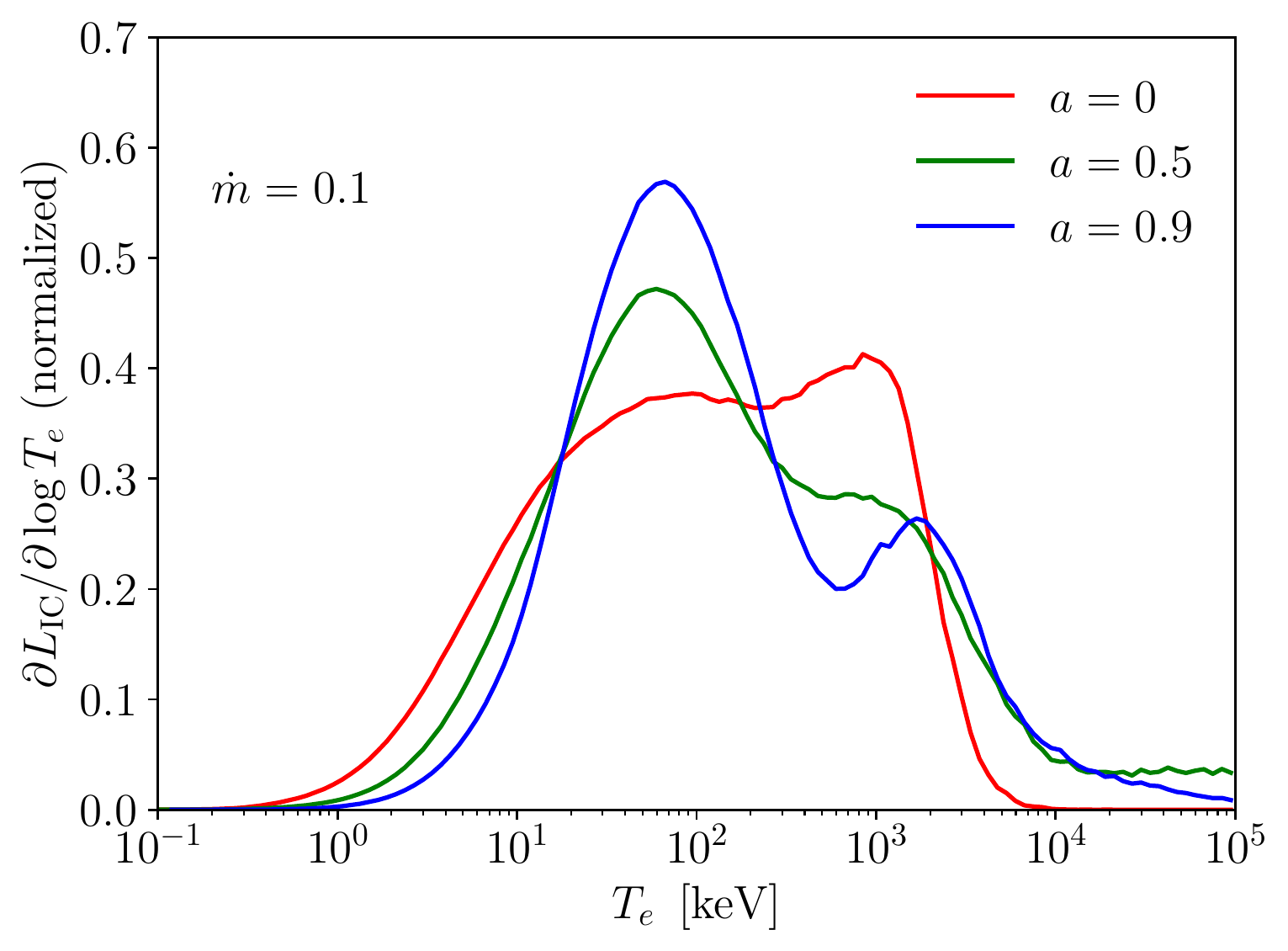}
\caption{For each of the six simulations as indicated, the curves represent the relative fraction of the total coronal cooling at a given electron temperature. These \textsc{harm3d}-calculated cooling rate values are integrated over the coronal volume and averaged over the entire $1000M$ run. All six curves are normalized so that the integrals with respect to $\log T_e$ are unity, for ease of comparison.
\label{fig:dLdT}}
\end{figure}

\begin{figure}
\plottwo{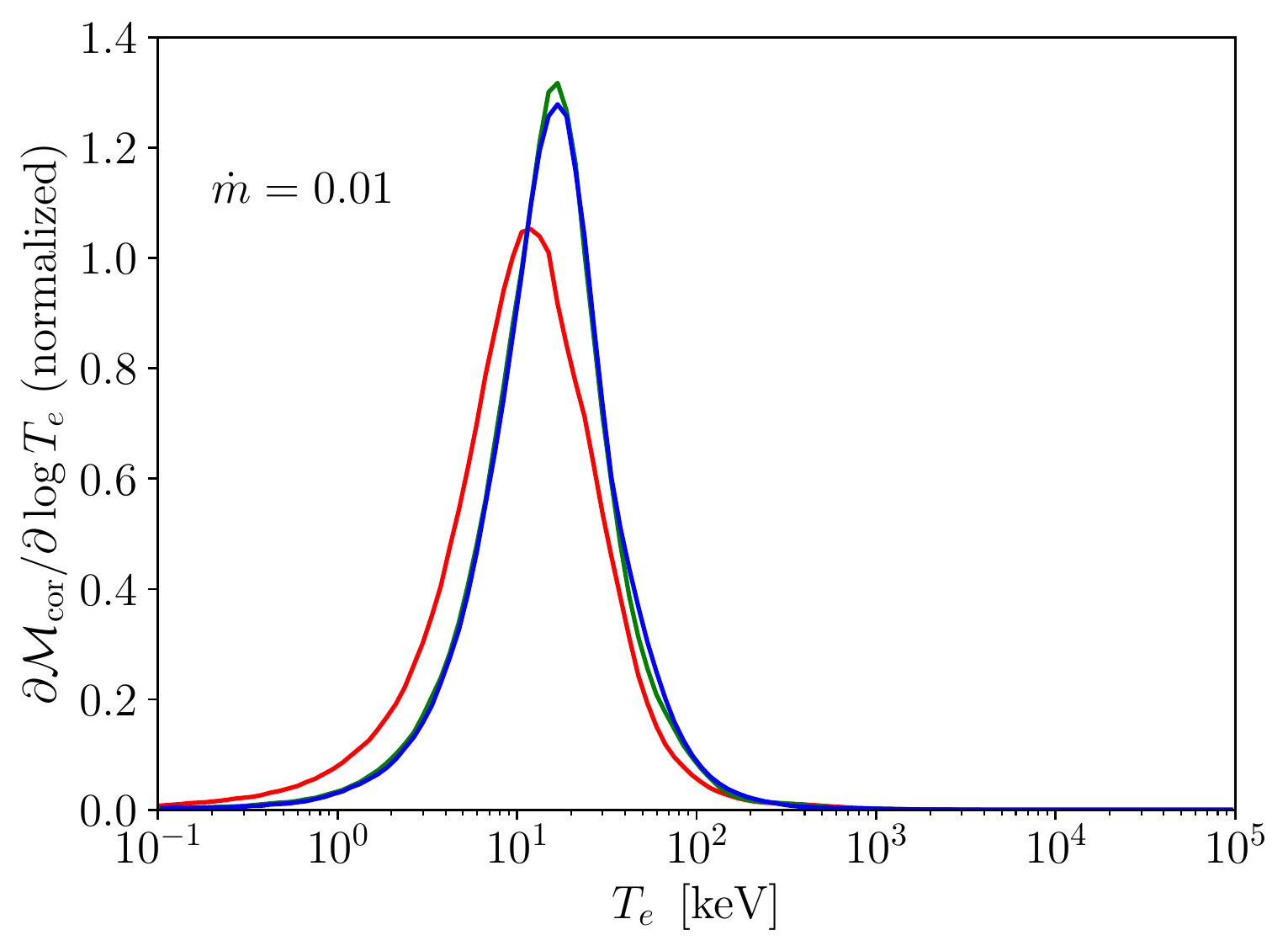}{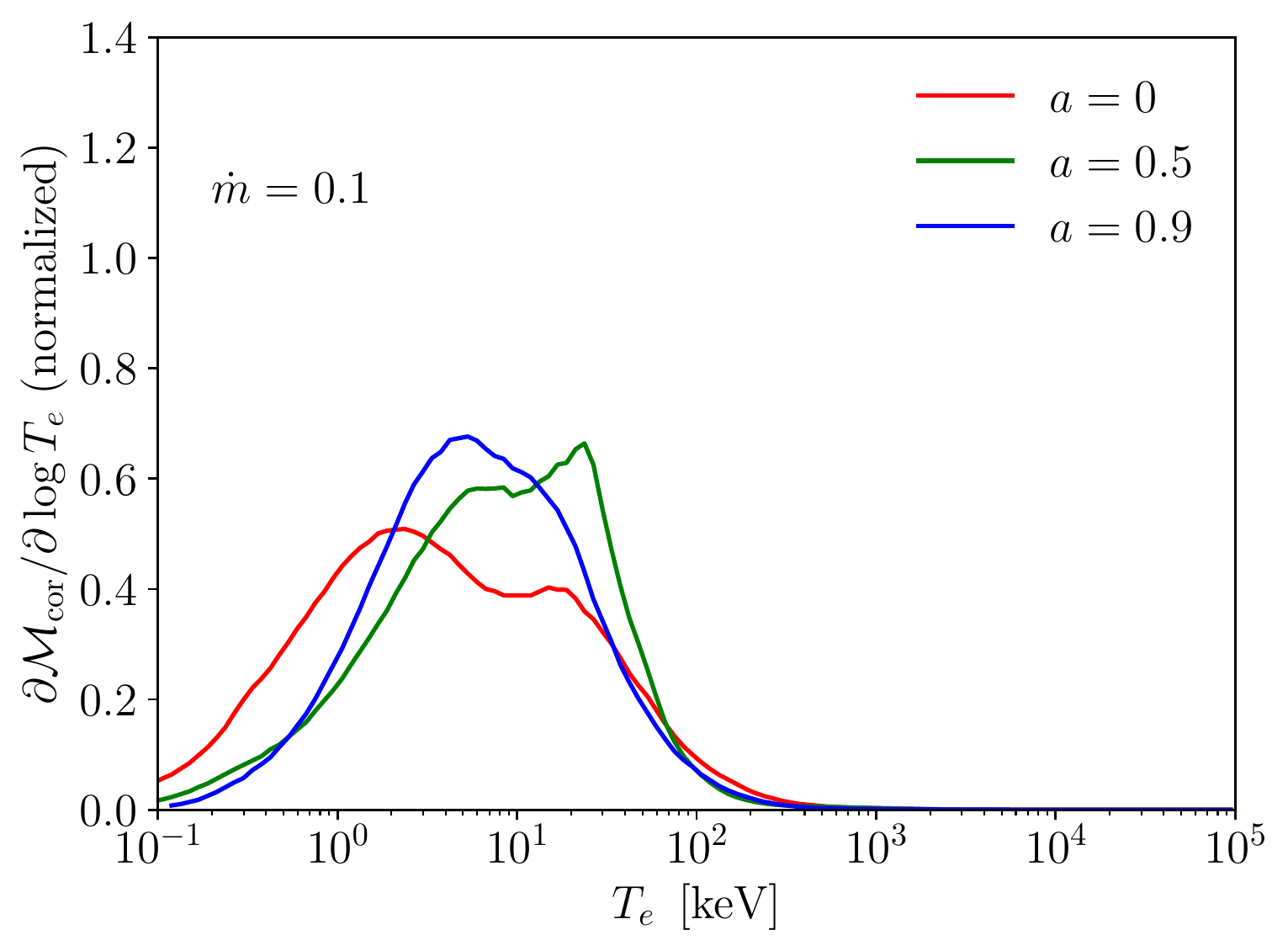}
\caption{Like Figure \ref{fig:dLdT}, but the curves indicate the relative fraction of the total coronal mass at a given electron temperature. Note that while a substantial fraction of the total cooling budget is due to gas hotter than 1 MeV, it accounts for essentially none of the corona's \emph{mass}.
\label{fig:dMdT}}
\end{figure}

The unphysical thermodynamics within the jet region, however, \emph{do not} affect the X-ray spectra we calculate from these simulations. In Figure \ref{fig:dLdT} we show how the corona's total cooling budget is distributed among gas of different temperature; likewise, we show in Figure \ref{fig:dMdT} how the corona's mass varies in temperature. The general trend is a greater portion of the coronal cooling and material at higher temperatures for the higher spin runs---though the $\dot{m} = 0.1$ simulation results display a more complex structure relative to the $\dot{m} = 0.01$ results. Note that while a fair share of the total coronal cooling is due to gas above 1 MeV, this same gas accounts for essentially none of the coronal mass budget. As explained in \citet{kin20a}, $L_\mathrm{IC} \propto \rho \Theta_e^2$ for $\Theta_e \gg 1$, where $\Theta_e$ is the dimensionless electron temperature ($\Theta_e = k_B T_e/m_e c^2$; $m_e c^2 = 511$ keV); thus for the super hot gas which resides  in the evacuated polar regions, the inverse Compton cooling rate (when nonzero) is often extreme, even when the density is very low.

\begin{figure}
\plottwo{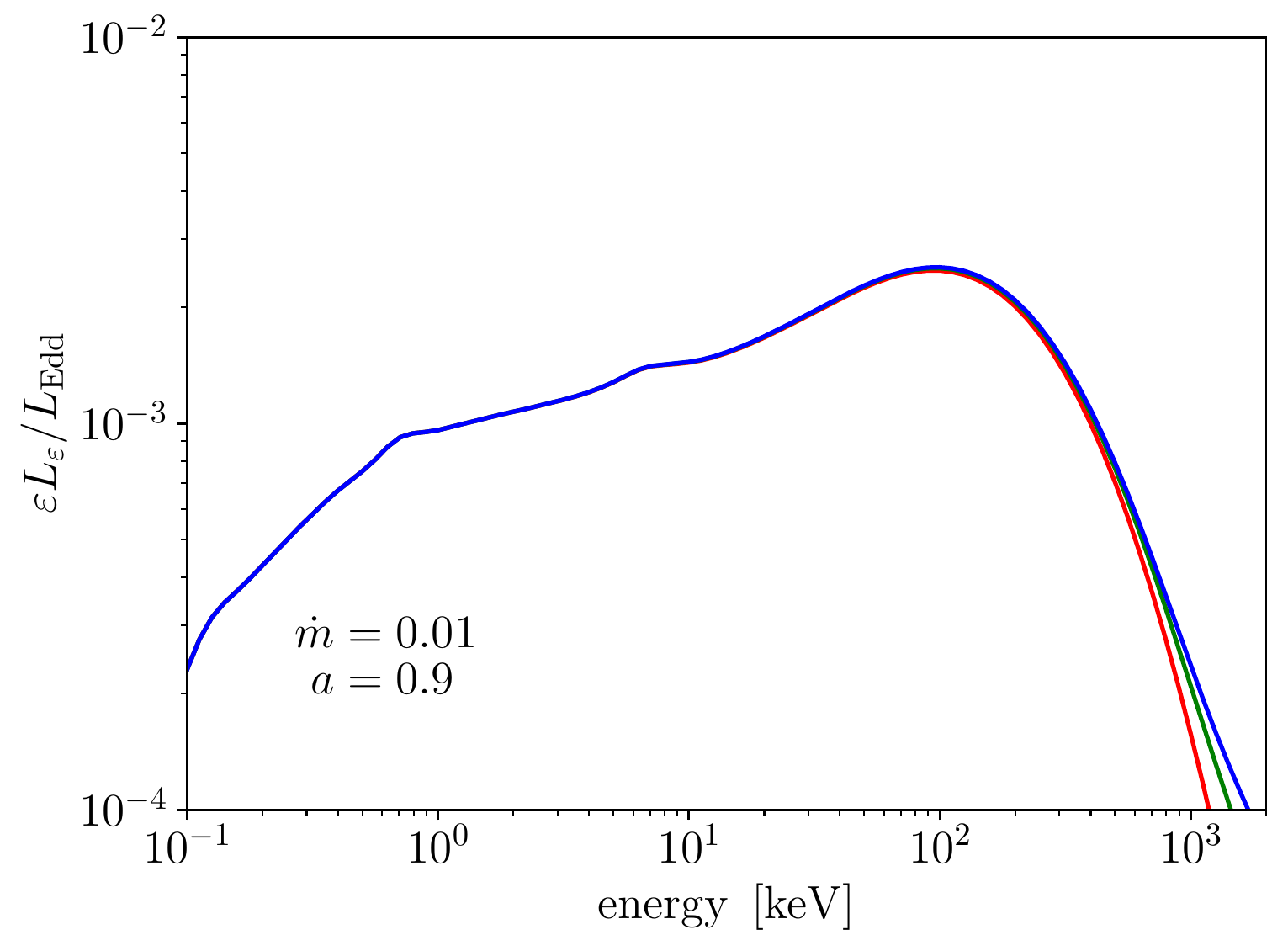}{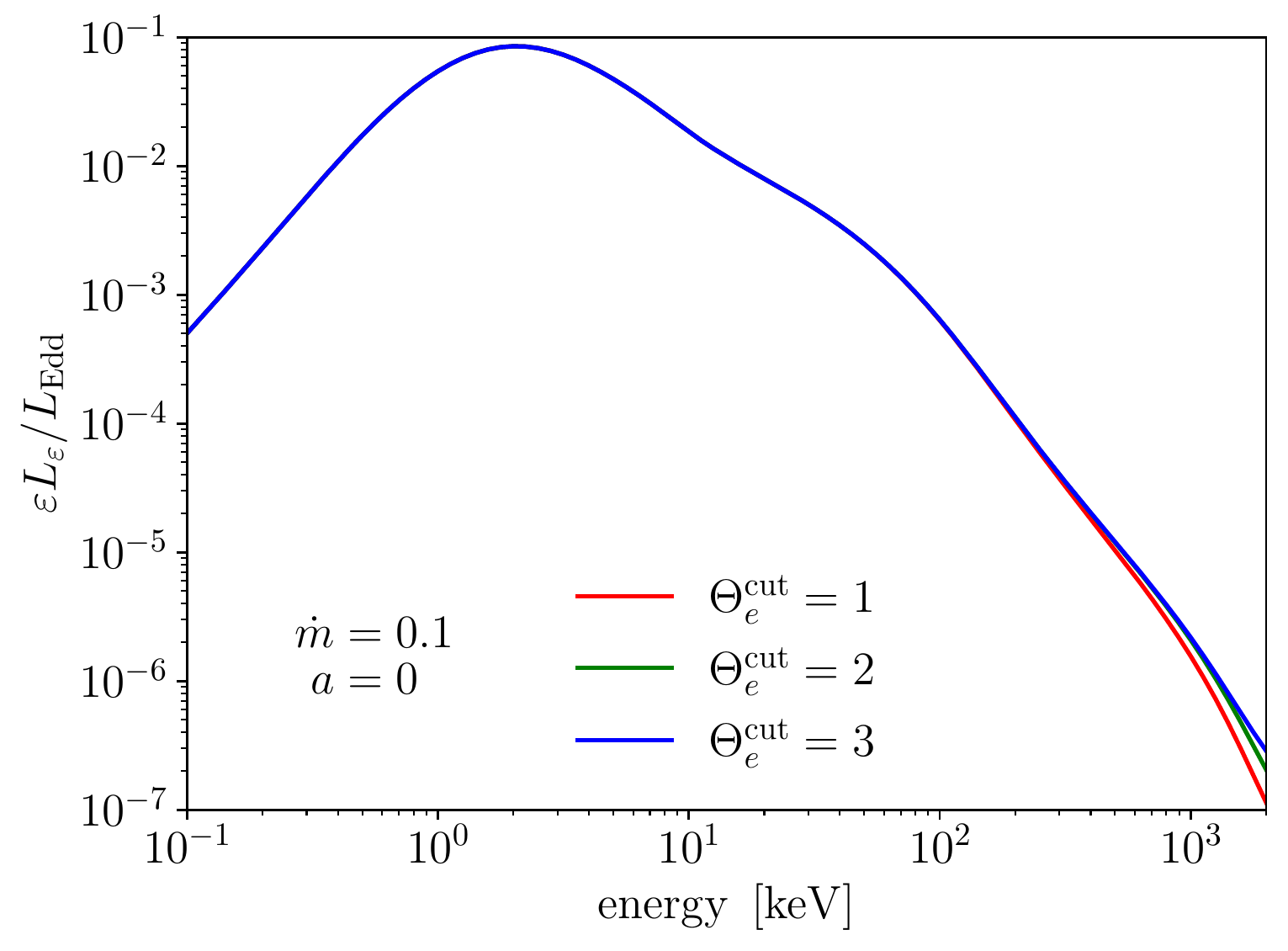}
\caption{The post-processed spectral luminosity (measured at infinity) for the two simulations indicated. For each, the spectra are shown after applying three different values of $\Theta_e^\mathrm{cut}$ in \textsc{pandurata}'s ray-tracing calculation, as described in the text. The value chosen for $\Theta_e^\mathrm{cut}$ affects the spectra only above 200 keV.
\label{fig:theta_cut}}
\end{figure}

In the next section we will discuss X-ray spectra at length---here we briefly explore how the anomalously hot polar gas affects spectral predictions. In Figure \ref{fig:theta_cut} we show, for two of the six simulations, the results of post-processing the $t = +1000M$ snapshots with the \textsc{pandurata}+\textsc{ptransx} pipeline described above. After its coronal $T_e$ map is determined, a value $\Theta_e^\mathrm{cut}$ is imposed before the final ray-tracing operation: inverse Compton scattering events are not simulated in cells with $k_B T_e/m_e c^2 > \Theta_e^\mathrm{cut}$. As we see, increasing $\Theta_e^\mathrm{cut}$---i.e., allowing \textsc{pandurata} to probe hotter gas---affects only the very high energy portion of the spectrum. Because this hot gas is so diffuse, \textsc{pandurata} \emph{under}-samples it even if it more than sufficiently samples the cooler, nearer to the disk regions of the corona. Increasing the number of photon packets by a factor of 20 leads to an observed spectrum that is indistinguishable from one constructed with the standard number all the way from 0.1 keV to 1 MeV; even at 2 MeV, the departure is only 10\%.

In practice, extraordinarily rare but large amplification scattering events can prevent \textsc{pandurata} from achieving a global solution---this is because, with the current code structure, it is not computationally feasible to simulate enough scattering events in the polar regions to reliably reconstruct a meaningful (Monte Carlo) cooling rate to then match to the (analytic) \textsc{harm3d} cooling rate. A biasing/splitting procedure could be employed to better sample these rare events---a method specialized for relativistic Compton scattering is discussed in \citet{kin21a}. However, because we are confident that the spectrum below 200 keV is unaffected by these rare events, we proceed by simply applying the no-scattering rule for any cell with $\Theta_e > 1$.

\begin{figure}
\plotone{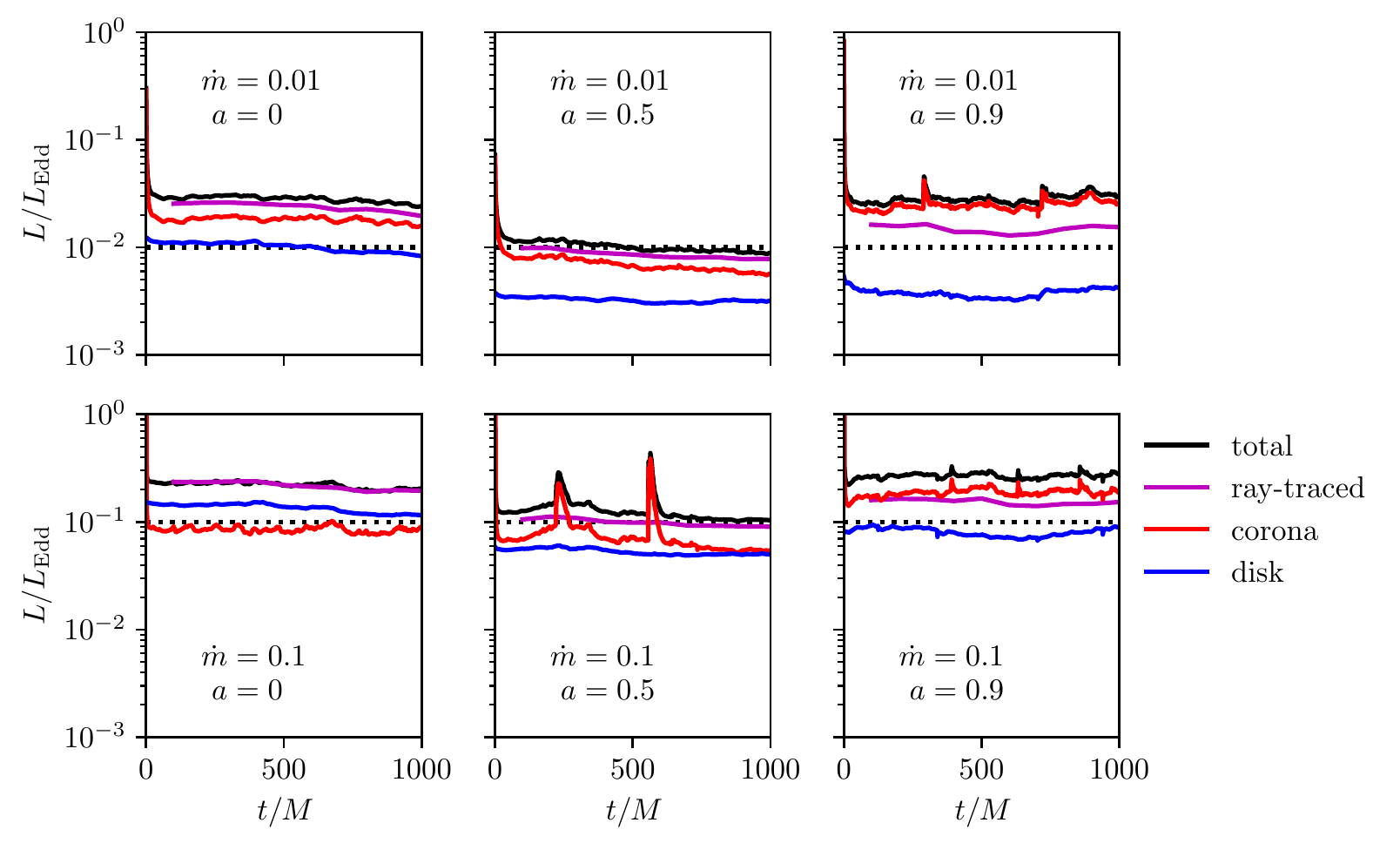}
\caption{The ray-traced luminosity, as well as the volume-integrated cooling rates, expressed in ratio to the Eddington luminosity, as functions of time after switching on the IC cooling function. The dotted lines indicate the nominal luminosity expected for the choice of accretion rate.
\label{fig:lum_grid}}
\end{figure}

In Figure \ref{fig:lum_grid}, we show the ray-traced luminosity and volume-integrated cooling rates as functions of time for each of the six simulations. Note that the corona is relatively more luminous than the disk as the spin increases, and relatively less so at higher accretion rate. This trend is summarized by the time-averaged $L_\mathrm{disk}/L_\mathrm{tot}$ values listed in Table 1. The higher spin runs tend to have more variable cooling rates, especially along the jet cone boundary. The prominent spikes in the $\dot{m} = 0.1$, $a = 0.5$ run, as well as the more numerous smaller spikes in the $a = 0.9$ runs, are due to very hot material falling below the magnetization threshold for evolution via the entropy equation and therefore rapidly cooling down following a sustained period of zero cooling. Figure \ref{fig:lum_grid} also shows the ray-traced value of the total luminosity, i.e., at what luminosity distant observers would measure the system. The ray-traced luminosity is always less than the volume-integrated value because the former accounts for gravitational redshift, capture of photons by the black hole, and---especially so for the higher spin runs---an increasing fraction of the total cooling occurring in the (deliberately or otherwise) inadequately sampled hot, diffuse polar region. The ray-tracing is performed only every $100M$ (whereas the three other curves in Figure \ref{fig:lum_grid} are sampled every $1M$) starting at $t = +100M$, so these curves appear artificially less variable. The ``spiking'' behavior typically occurs in gas at sufficiently high temperature that it does not affect the ray-traced X-ray luminosity.

\begin{figure}
\plotone{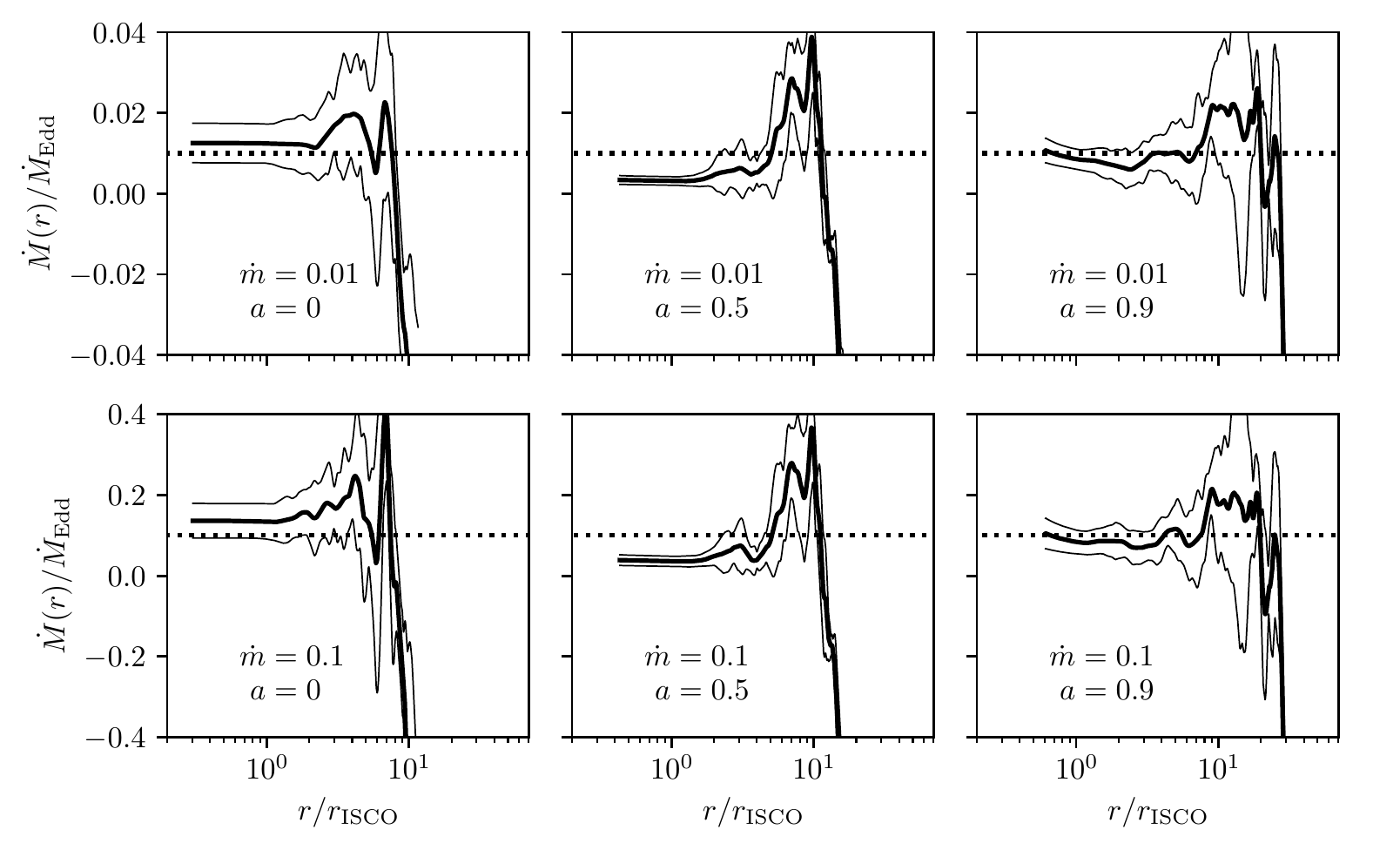}
\caption{The time-averaged net mass inflow rate, expressed in ratio to the Eddington accretion rate (assuming $\eta = \eta_\mathrm{NT}$). The thin lines above and below indicate plus-or-minus one standard deviation from the time-averaged value. The dotted lines indicate the nominal net mass inflow rate expected for this choice of accretion rate scaling (again, assuming $\eta = \eta_\mathrm{NT}$). The $x$-axes are expressed in multiples of the ISCO radius: for $a = 0$, 0.5, and 0.9, $r_\mathrm{ISCO}/M = 6$, 4.23, and 2.32, respectively.
\label{fig:inflow_grid}}
\end{figure}

Figure \ref{fig:inflow_grid} shows the time-averaged, net mass inflow rate as a function of radius (in units of the ISCO radius) for each of the six simulations. Time-averaged mass inflow equilibrium is established out to at least a few multiples of the ISCO radius for all runs.

The time-averaged values for ray-traced luminosity, the total volume-integrated cooling rate, and the mass inflow rate (measured at the event horizon) allow us to calculate post hoc values for the radiative efficiency, listed in Table 1. The volume-integrated cooling rate value is reported down the ``$\eta$'' column; the values computed using the ray-traced luminosity are labeled ``$\eta_\mathrm{rt}$,'' and the classical geodesic thin disk efficiency given by \citet{nov73a} is ``$\eta_\mathrm{NT}$.'' Note that the $\eta$ values tend to be a factor $\sim$ 1.5--3 times the $\eta_\mathrm{NT}$ values, while the $\eta_\mathrm{rt}$ values are not quite so large (but still always greater than $\eta_\mathrm{NT}$). The values for $\eta_\mathrm{rt}$ are lower than the direct volume-integrated $\eta$ values because, as discussed above, there is a substantial fraction of the total cooling budget which is not reflected in the ray-tracing procedure, especially for high spin, as the low density in the jet cone inhibits adequate Monte Carlo sampling of the inverse Compton power. A secondary but much smaller effect is the capture of photons by the black hole, which is accounted for in determining $\eta_\mathrm{rt}$, but not in $\eta$ or $\eta_\mathrm{NT}$ [though see \citet{sch16a}]. Thus $\eta_\mathrm{rt}$ is in this sense a lower bound for the predicted radiative efficiency.

In addition, the inferred radiative efficiency values are a strong increasing function of the spin (like $\eta_\mathrm{NT}$) but vary little with the nominal accretion rate, as is expected since the target-temperature cooling function, the IC cooling function, and the mass inflow rate all depend linearly on density. The distribution of cooling between corona and disk, however, does depend on the nominal accretion rate choice---as we see below, this division significantly impacts the resulting X-ray continuum spectra.

\begin{deluxetable}{ll|cccc}
\tablecaption{Summary of Key Simulation Results:}
\tablenum{1}
\tablehead{\colhead{$\dot{m}$} & \colhead{$a$} & \colhead{$L_\mathrm{disk}/L_\mathrm{tot}$} & \colhead{$\eta$} & \colhead{$\eta_\mathrm{rt}$} & \colhead{$\eta_\mathrm{NT}$}}
\startdata
0.01 & 0 & 0.357 & 0.1284 & 0.1086 & 0.0572 \\
0.01 & 0.5 & 0.319 & 0.2460 & 0.2096 & 0.0821 \\
0.01 & 0.9 & 0.131 & 0.4164 & 0.2170 & 0.1558 \\
0.1 & 0 & 0.612 & 0.0917 & 0.0910 & 0.0572 \\
0.1 & 0.5 & 0.407 & 0.2739 & 0.2094 & 0.0821 \\
0.1 & 0.9 & 0.299 & 0.4010 & 0.2292 & 0.1558 \\
\enddata
\tablecomments{For the six simulations performed, characterized by their accretion rate ($\dot{m}$) and spin ($a$), we report time-averaged values for the fraction of total luminosity in the disk ($L_\mathrm{disk}/L_\mathrm{tot}$), the inferred radiative efficiency using the volume-integrated cooling rate ($\eta$), and the inferred radiative efficiency using the ray-traced luminosity ($\eta_\mathrm{rt}$). For comparison, the Novikov-Thorne radiative efficiency for each spin value ($\eta_\mathrm{NT}$) is also listed. These values are calculated for a $10 M_\odot$ black hole.}
\end{deluxetable}

\clearpage

\subsection{Predicted Spectra}

\begin{figure}
\plotone{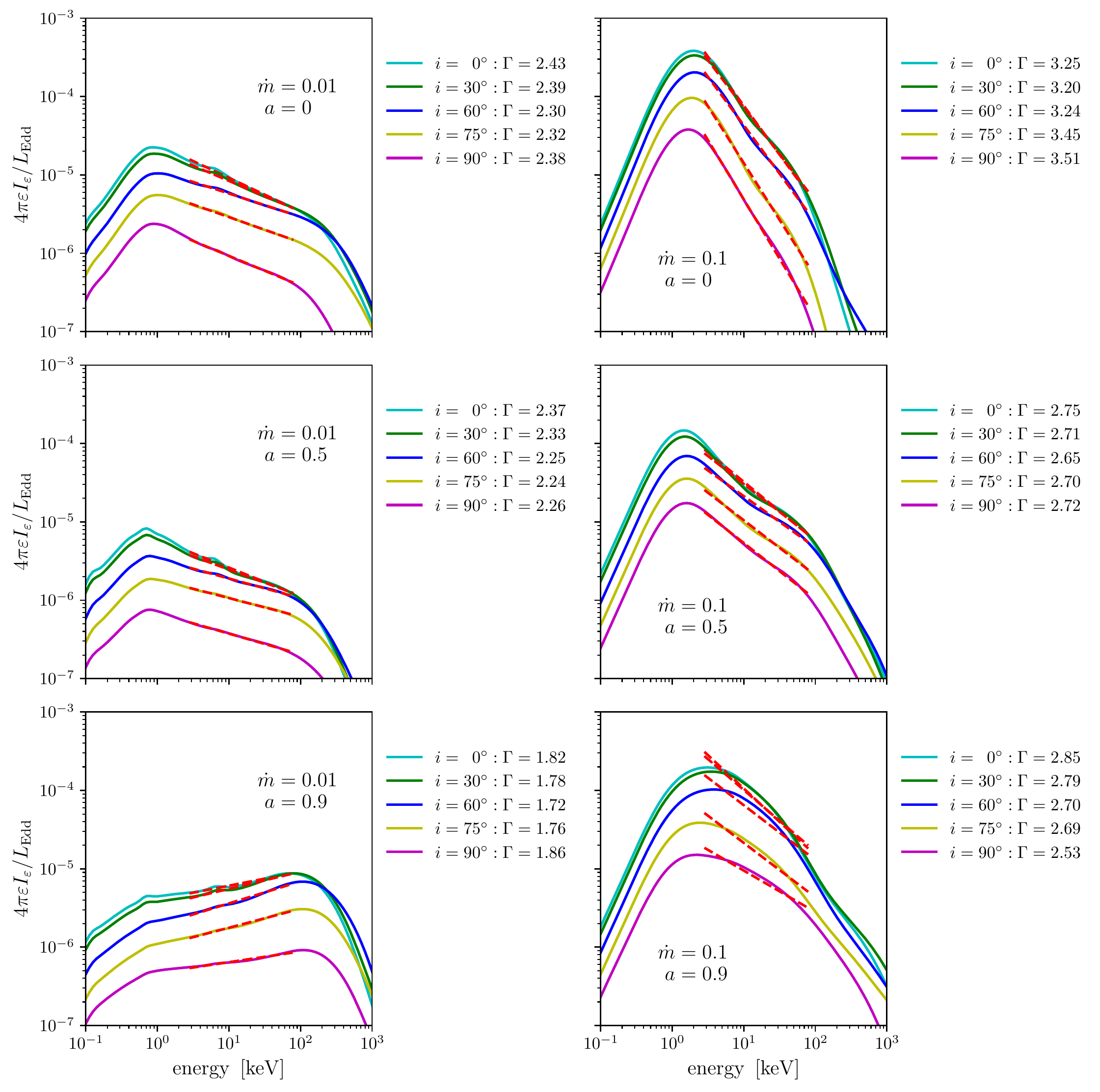}
\caption{The spectral intensity as viewed by distant observers at a sampling of inclination angles, for the $t = +1000M$ snapshots from each of the six simulations as indicated. The dashed red lines are power law fits to the NuSTAR band for each spectrum, with associated photon index reported.
\label{fig:inc_grid}}
\end{figure}

In Figure \ref{fig:inc_grid}, we show the spectral intensity for a series of observers at infinity (characterized by their viewing angle $i$), as calculated by our \textsc{pandurata}+\textsc{ptransx} post-processing pipeline for snapshots of each of the six simulations $1000M$ after switching on the IC coronal cooling function. Superimposed on each spectrum is a power law fit to the NuSTAR X-ray band, 3--79 keV, with associated photon index $\Gamma$. The overall trend is immediately apparent: the spectrum \emph{hardens} with increasing spin, and \emph{softens} with increased accretion rate. The hardening with spin is due to the increased coronal temperature at higher spins (see Figure \ref{fig:harm_Te_grid}), while the softening with accretion rate is due to the larger share of total cooling occurring in the disk (and therefore radiating thermally) at higher accretion rate (see Table 1).

\begin{figure}
\plotone{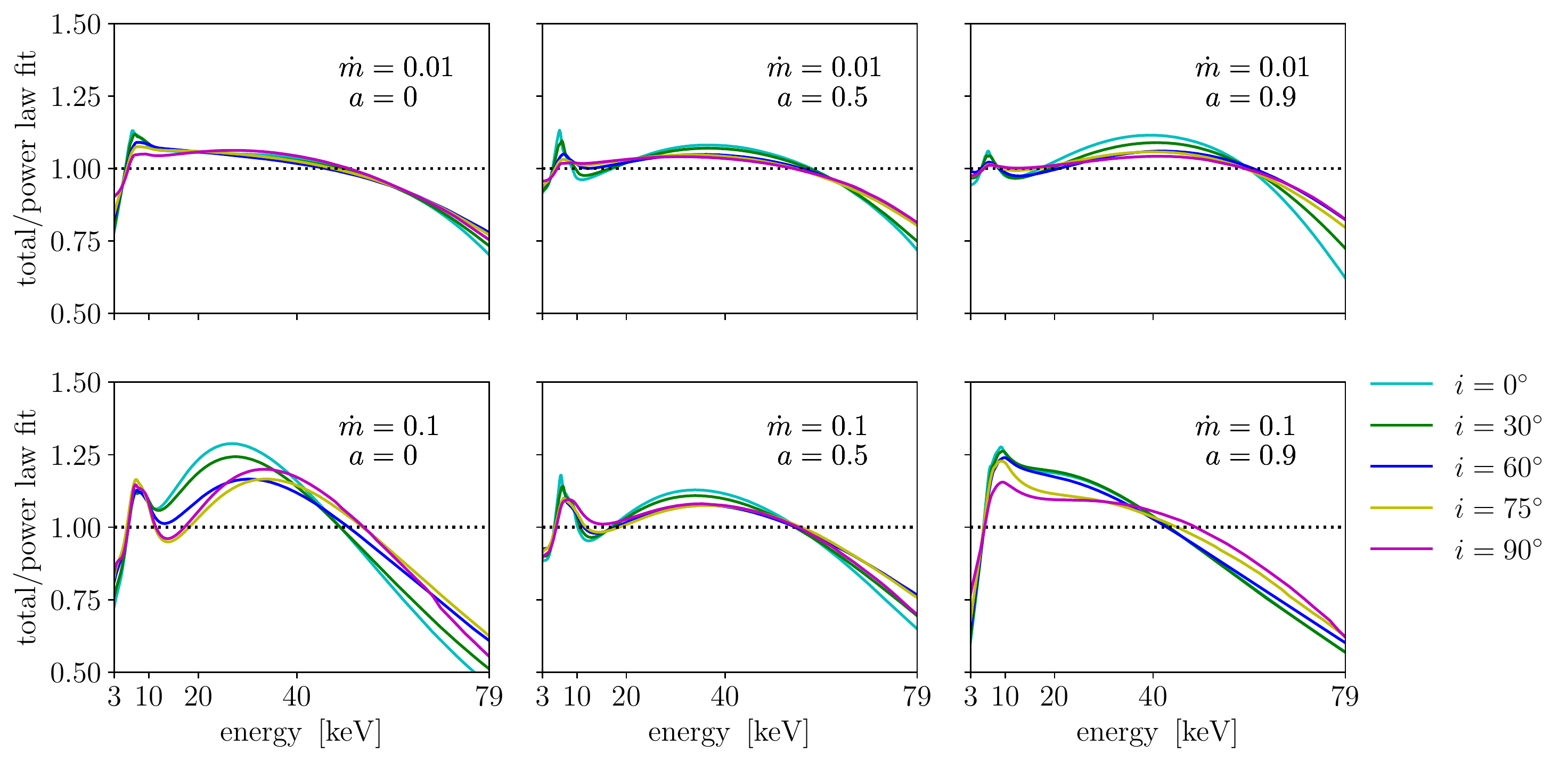}
\caption{The distant observer spectra divided by their power law fits at a sampling of viewing angles. The feature at $\sim$ 6 keV is the Fe K$\alpha$ emission line, while the broader deviation at 20--50 keV is the ``Compton bump.''
\label{fig:pl_ratio}}
\end{figure}

Not surprisingly, the smaller solid angle the disk subtends for the more inclined viewers results in a diminished overall flux as $i$ increases from $0^\circ$ to $90^\circ$. Also, the peak hardness typically occurs for viewers inclined at $\sim 60^\circ$; this is due to the larger scattering optical depth traversed by photons emerging more nearly parallel to the plane of the disk (relative to those that escape through, e.g., the extremely optically thin jet cone). This trend is less well-established for the $\dot{m} = 0.1$ runs, in part because these spectra more substantially deviate from power laws in the NuSTAR band. This deviation is depicted in Figure \ref{fig:pl_ratio} by dividing the distant observer spectra by their power law fits.



The prominent Fe K$\alpha$ emission feature at $\sim 6$ keV we discuss further below. Note that the ``Compton bump'' between 20--50 keV is enhanced for the higher accretion rate runs. This feature arises from high energy (already Compton upscattered) corona photons undergoing Compton downscattering when they reflect off the cool disk, and is enhanced further by the Fe K-edge photoionization absorption opacity above $\sim 10$ keV---the absorption of Compton downscattering photons by Fe forms the boundary between the Fe K$\alpha$ emission feature and the Compton bump.

The elevated photosphere surfaces of the higher accretion rate runs (recall Figure \ref{fig:harm_rho_grid}) increase the likelihood of coronal photons colliding with the disk surface while traversing the corona; in addition, the softer incident spectra striking the cooler outer disk layers results in a greater fraction of the disk-incident photons undergoing K-edge absorption. Both of these effects contribute to the enhanced Compton bumps present in the $\dot{m} = 0.1$ spectra. Note that the $\dot{m} = 0.1$, $a = 0.9$ spectra are so poorly-described by a power law in the NuSTAR band that the bump is not as well-defined in this view.

\begin{figure}
\plotone{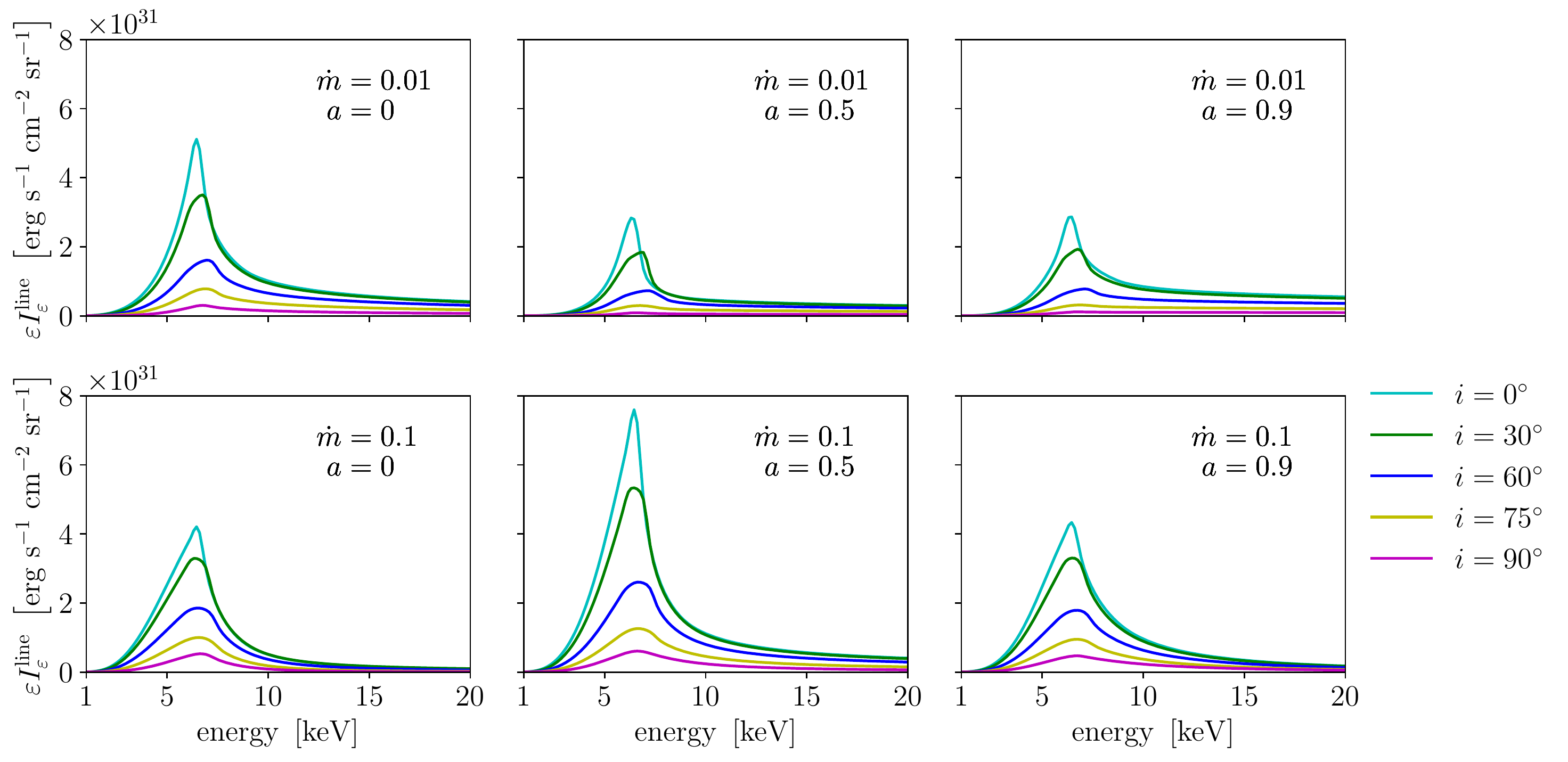}
\caption{The ``true'' continuum-subtracted Fe K$\alpha$ line profiles, for a sampling of viewer inclination angles. Only those photons whose physical origin is a bound-bound transition in the range 6.3--7.0 keV are included.
\label{fig:line_flux}}
\end{figure}

Figure \ref{fig:line_flux} shows the ``true'' Fe K$\alpha$ emission line feature. We define ``true'' Fe K$\alpha$ emission as those line photons originating from a bound-bound transition with photon energies in the range 6.3--7.0 keV; by tagging these photons at creation in \textsc{ptransx} and tracking them through \textsc{pandurata}, we are able to identify the ``line'' and ``continuum'' components of a predicted spectrum by their physical origin, without resorting to a fitting procedure for isolating the continuum. The corresponding ``true'' equivalent width of these K$\alpha$ emission features all lie in the range 100--500 eV, which are in line with typical observed strengths.

The line profiles shown in Figure \ref{fig:line_flux} are analogous to a typical continuum-subtracted representation, but with the continuum known exactly (Figure \ref{fig:pl_ratio} is more akin to what is usually done in practice, i.e., a fit to the underlying continuum). Though the line profiles we predict are relativistically-broadened, the occasionally-observed ``double-horned'' feature (due to Doppler shifting of advancing and receding material in the disk) is essentially completely smoothed over by Compton scattering in the corona.


\begin{figure}
\plottwo{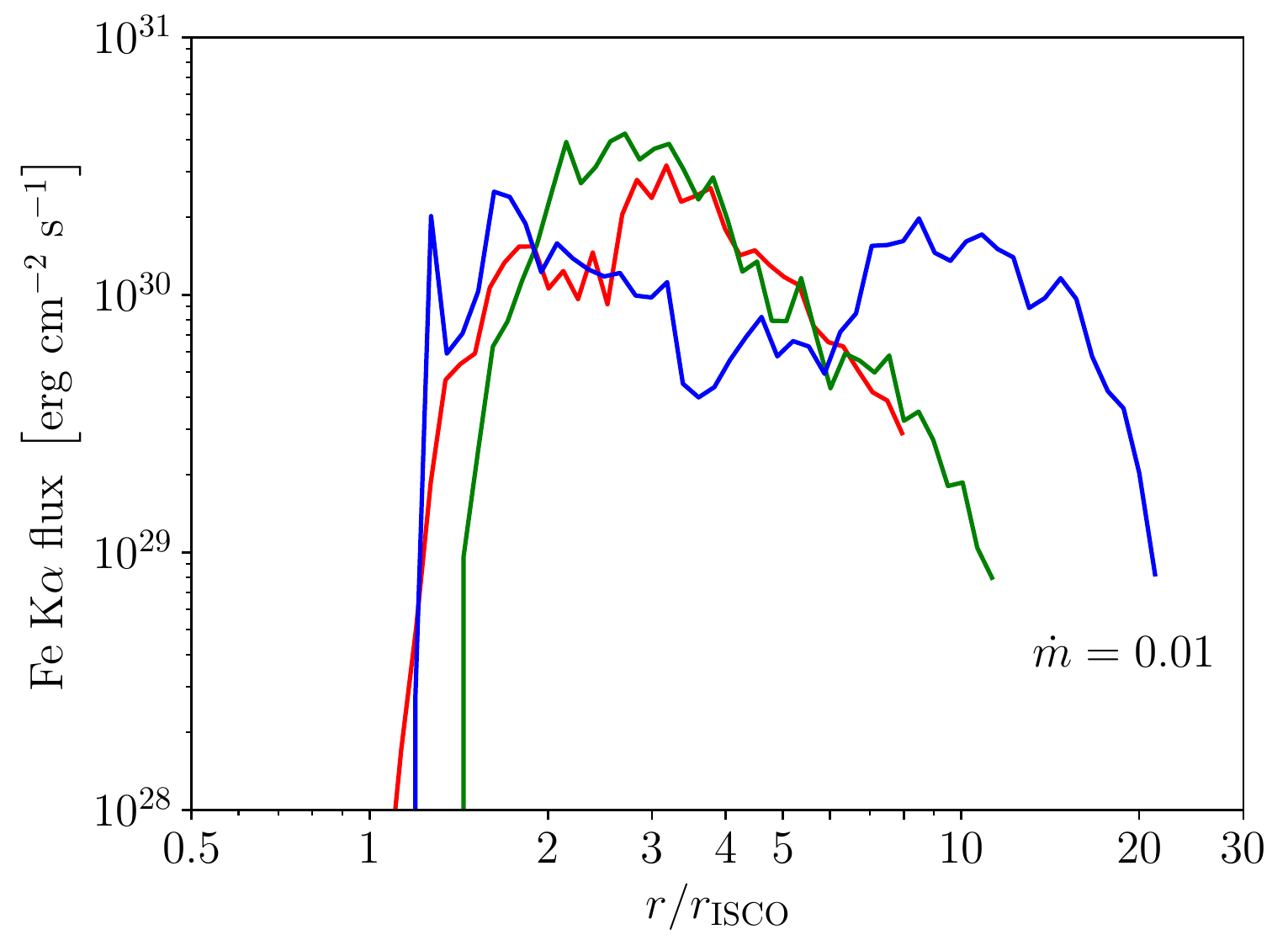}{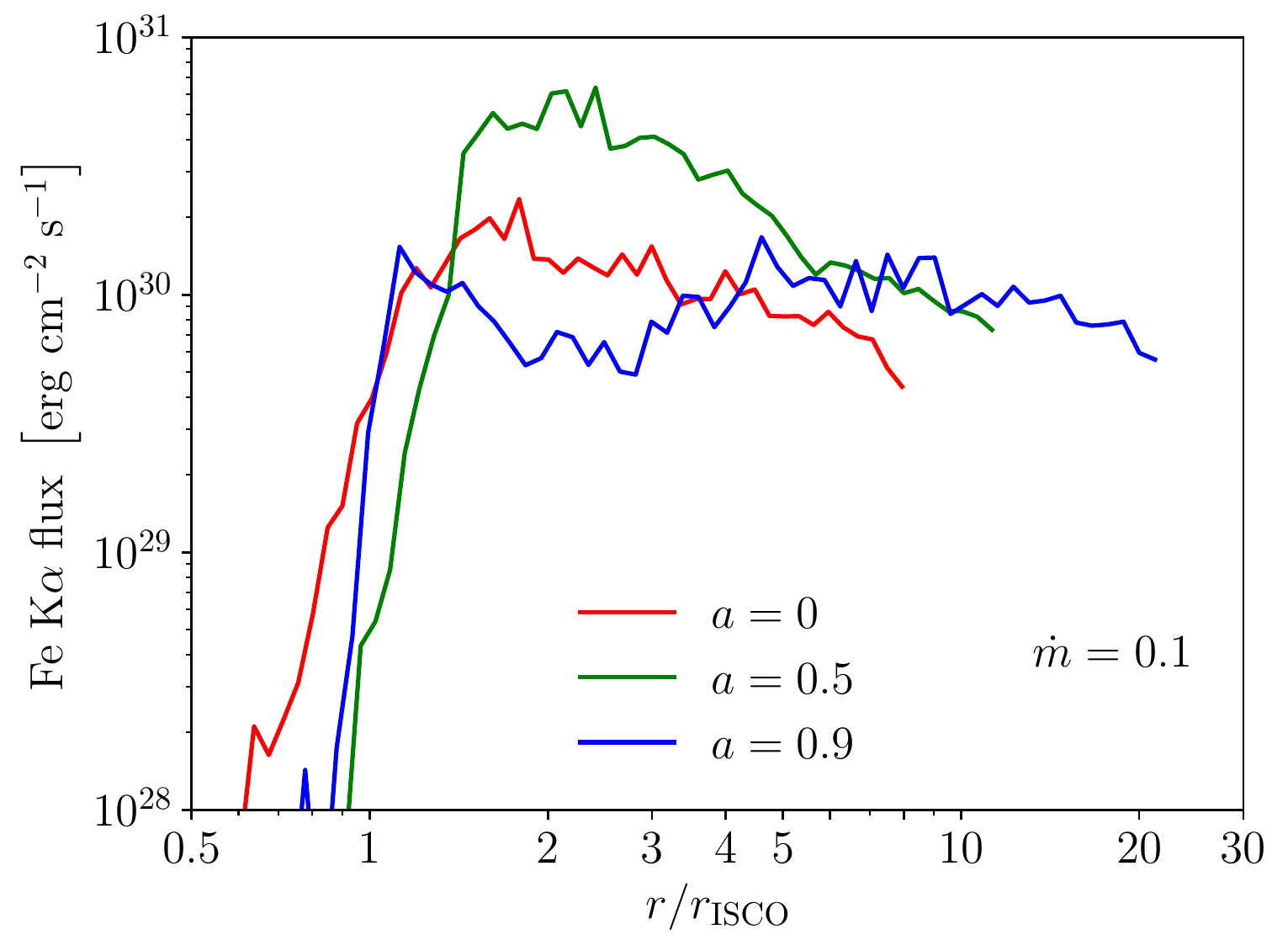}
\caption{The $\phi$-, top- and bottom-averaged Fe K$\alpha$ surface brightness for each of the six simulations at $t = +1000M$. The $x$-axis is expressed in multiples of the ISCO radius: for $a = 0$, 0.5, and 0.9, $r_\mathrm{ISCO}/M = 6$, 4.23, and 2.32, respectively. While K$\alpha$ production terminates \emph{near} the ISCO radius, it does not do so precisely \emph{at} the ISCO, nor does it decline steeply with increasing radius as is usually assumed with a ``lamppost''-style corona model. Moreover, the location and sharpness of the inner K$\alpha$ cut-off depend on accretion rate and spin.
\label{fig:feka_rad}}
\end{figure}

Unlike the continuum, the Fe K$\alpha$ emission profiles we calculate do not show strong, immediately apparent trends. The red wing in particular---often the primary driver for spin measurements in X-ray reflection spectroscopy techniques---is nearly unchanged as a function of spin when normalized for total K$\alpha$ emission. Figure \ref{fig:feka_rad} shows how the disk surface emission of Fe K$\alpha$ varies with multiples of the ISCO radius: a typical emission model assumes a power law-decaying surface brightness profile with a sharp inner cutoff at the ISCO radius. While our predicted surface brightness profiles do in fact terminate near (but not exactly at) the ISCO radius (and the termination is more abrupt at $\dot{m} = 0.01$ than for $\dot{m} = 0.1$), they do not decline rapidly with increasing radius---rather, they vary haphazardly and only gently decline with radius (except for the $a = 0.9$ curves, which are roughly flat), reflecting the turbulent accretion flow and the spatially-extended corona. The most substantial trends concerning the Fe K$\alpha$ emission are: its equivalent width and peak contrast tend to increase with accretion rate, and the \emph{blue} wing tends to strengthen at higher spin.

\begin{figure}
\epsscale{0.6}
\plotone{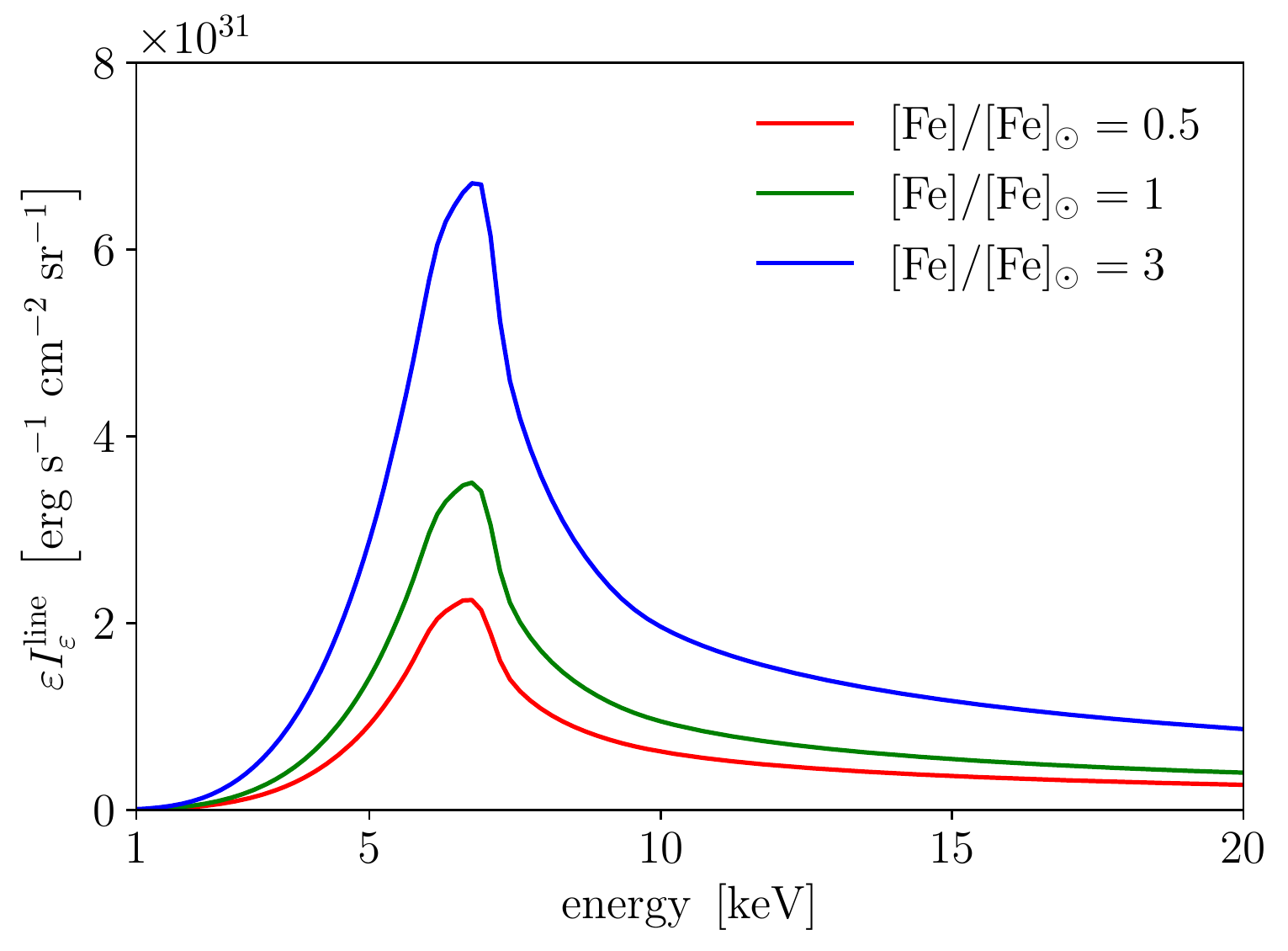}
\caption{The ``true'' (continuum-subtracted) Fe K$\alpha$ line profiles for the same simulation ($\dot{m} = 0.01$, $a = 0$), at the same viewing angle ($i = 30^\circ$), for three different values of the Fe abundance. Note that the strength of the feature increases sub-linearly with the Fe abundance; scaling to the overall K$\alpha$ flux, the line profile \emph{shapes} are essentially identical.
\label{fig:line_abund}}
\end{figure}

Thus far we have used a solar elemental composition. In Figure \ref{fig:line_abund}, we show the K$\alpha$ line profiles which result from post-processing the $t = +1000M$ snapshot of the $\dot{m} = 0.01$, $a = 0$ simulation, but with the Fe abundance at half, equal to, and three times the solar value. As one might expect, the Fe K$\alpha$ line strength increases with Fe abundance, though sub-linearly. Above some abundance the K$\alpha$ strength would saturate, as the availability of K-edge photons (not Fe atoms) would become the limiting factor for K$\alpha$ production. The underlying continuum (not shown) does change slightly with Fe abundance, as the K-edge opacity and corresponding photoionization heating affects the temperature structure of the disk.



\begin{figure}
\epsscale{0.6}
\plotone{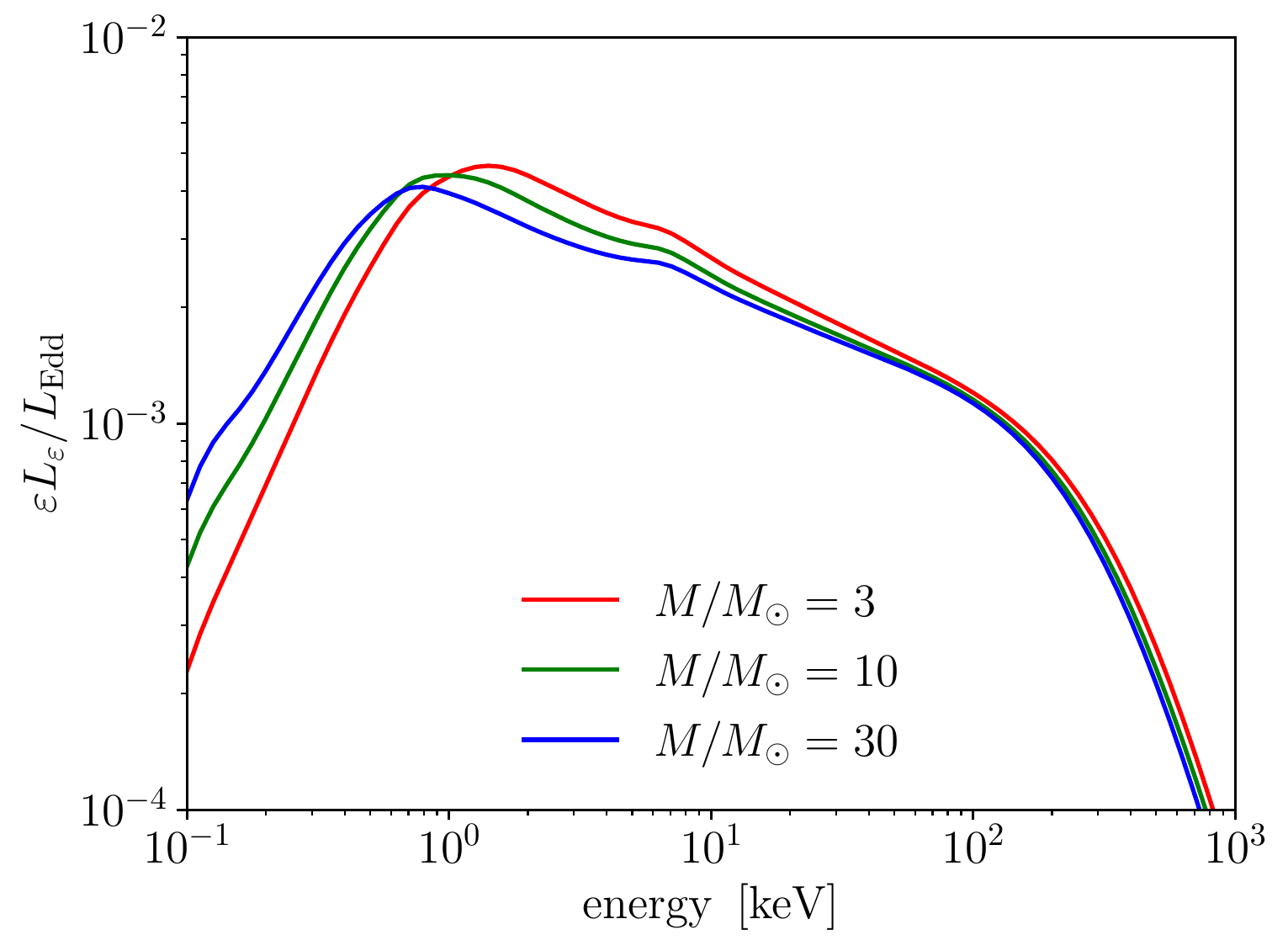}
\caption{The spectral luminosity for a distant observer, for the $t = +1000M$ snapshot of the $\dot{m} = 0.01$, $a = 0$ simulation, scaled to the three indicated values of the central black hole mass.
\label{fig:spec_mass}}
\end{figure}

So far we have always set $M = 10 M_\odot$. Figure \ref{fig:spec_mass} shows the post-processed spectrum of the $\dot{m} = 0.01$, $a = 0$ run, with central black hole masses equal to 3, 10, and 30 $M_\odot$. On the scale shown (relative to the Eddington luminosity, which also scales with mass), the change is minor: as the mass increases, the thermal peak recedes---as expected, as $T_\mathrm{disk} \propto M^{-1/4}$---while the power law component hardens slightly due to the lower and redder thermal peak. The effect of central black hole mass on the Fe K$\alpha$ feature is very small: the Fe K$\alpha$ equivalent width changes by only $\lesssim 10\%$ over the given mass range, and not in a uniform way; additionally, the continuum-subtracted line shape remains nearly unchanged after controlling for the overall linear scaling of flux with mass.

\clearpage

\section{Discussion}

As emphasized in the introduction and also in our preceding methods and ``proof of concept'' papers, our 3D GRMHD simulation-derived spectral predictions are the most directly connected to the underlying physics of any X-ray binary spectral model available. The disk (thermal) and coronal (power law) contributions to the continuum, the strength and shape of the Fe K$\alpha$ emission line, the magnitude of the Compton bump reflection feature, and location of the high energy rollover---all these features emerge naturally from the self-consistent treatment of the physical processes responsible, against a background accretion flow geometry provided by high resolution, MRI-resolving simulations. Because the parameters of our model all specify the actual properties of the black hole system---and the spectral properties follow suit without additional parameterization---we have made \emph{forward} predictions for how we expect these underlying properties to affect observations. Below we review the key results of this survey:

\begin{enumerate}

\item \emph{The Fe K$\alpha$ line profile shape is much more strongly affected by the accretion rate than by the spin.}

Contrary to the usual assumption in X-ray spectral modeling, the inner radial cutoff of K$\alpha$ emission we predict (for the range of accretion rates and spins we consider) is always outside the ISCO, though it does move inward (and terminates less sharply) for the higher accretion rate runs. The spatially-extended coronae our simulations develop illuminate the disks uniformly: the result is a radial K$\alpha$ surface brightness profile which, outside the cutoff, is fairly flat, in contrast to the usually-assumed steep profiles \citep{wil12a, dau13a} associated with lamppost geometries. Flat radial profiles result in relatively unchanging K$\alpha$ red wings with increasing spin as the fraction of the total K$\alpha$ photons which originate deep in the potential increases only a very small amount.

The K$\alpha$ profiles do, however, broaden and strengthen for the higher accretion rate runs. Consider the line profiles in Figure \ref{fig:line_flux}: the effect of raising the accretion rate of a non-spinning black hole is to enhance the red wing by a greater degree than does increasing the spin to 0.9. This is because the location of the inner cutoff is in fact determined not \emph{only} by the spin (through the ISCO radius), but also by the overall density scale as set by the accretion rate. An overall greater disk column density means more radiating Fe atoms at smaller radii, thereby pushing the inner cutoff farther inward.

As expected, the blue wing of the Fe K$\alpha$ line profile tends to strengthen (relative to the underlying continuum) at more edge-on views, due to relativistic beaming of K$\alpha$ photons from the rapidly rotating inner disk \citep{fab89a} and the corresponding enhanced electron scattering optical depth along more inclined lines of sight.

The Fe K$\alpha$ line strength (as measured by its equivalent width) increases sub-linearly with Fe abundance. For the example shown, raising the Fe abundance by a factor of 6 results in an equivalent width which grows only by a factor of 2.5. We note also that the K$\alpha$ line profile \emph{shape} is nearly identical after controlling for total K$\alpha$ photons seen by a given distant observer---the relative distribution of K$\alpha$ production over the disk surface is not substantially affected by the Fe abundance.

\

\item \emph{The overall radiative efficiency is at least 30--100\% greater than predicted by the Novikov-Thorne model.}

From analytic accretion disk theory, we expect an overall increase in the radiative efficiency at higher spin---we find the same trend in our simulations, though the radiative efficiency values we calculate are substantially greater than the analytic results (compare the $\eta_\mathrm{rt}$ values we infer from our calculations to the $\eta_\mathrm{NT}$ column in Table 1).

The classical radiative efficiency of the Novikov-Thorne model is based upon the assumption of zero stress at the ISCO radius, which automatically identifies the radiative efficiency with the binding energy of a particle on an ISCO orbit. The justification for this assumption has been challenged in many papers \citep{tho74a, kro99a, gam99a, nob09a, nob10a, ava16a} on the grounds that magnetic stresses are likely to be continuous across the ISCO. Based solely on these grounds, the actual efficiency has previously been estimated to be modestly elevated above NT \citep{nob09a, nob11a, sch16a} when the magnetic flux near the horizon is not too large, and possibly elevated rather more if it is very large [\citet{ava16a}, but see \citet{tei18a} for a contrary view]. The enhancement we find here is much larger than in this previous work, even though our methods are very close to those of \citet{nob09a}. We therefore attribute this change to the one new element relative to our previous work: the physical cooling function applied in the corona.

A similar result was reported for the non-spinning case in \citet{kin20a}---as explained there, the more physically realistic inverse Compton coronal cooling function turns out to be much more efficient at radiating away heat than was the old target-temperature formulation. The result is that more magnetic heating in the corona is radiated away as upscattered photons instead of being advected through the event horizon or carried out of the simulation volume. Importantly, this is a substantial increase in radiative efficiency above NT in the presence of only moderately-strong magnetic fields, in contrast to the ``magnetically arrested disk'' state typically associated with similar increases above NT \citep{ava16a}.

\item \emph{The power law component of the X-ray spectrum hardens with increasing spin, while the thermal component strengthens with increasing accretion rate. The Compton bump is enhanced by higher accretion rate and lower spin.}

These are the primary ways the X-ray continuum depends on the parameters varied. In our simulation-to-spectrum pipeline, the code units density converts to cgs units according to $\rho \propto \dot{m}/\eta_\mathrm{NT}$, where $\eta_\mathrm{NT}$ is an increasing function of the dimensionless spin (listed in Table 1). Even if we scaled the density using our inferred $\eta$ values instead, the trend with spin is similar. This scaling relationship of density with accretion rate is a central assumption in our calculation, though a roughly linear relationship is supported by recent simulations which include radiation forces \citep{jia19b}. For a fixed $\dot{m}$ value, the physical density scale decreases with increasing spin, resulting in photosphere surfaces which are closer to the midplane, and therefore contain a smaller portion of the (overall more efficiently cooling) simulation volume. For increasing $\dot{m}$ values at a fixed spin, the photospheres move farther from the midplane, and a greater portion of the (again increased) total cooling is captured by the disk. In this way, values for $\dot{m}$ and $a$ dictate the division between disk and corona and, thereby, the broad trends of the X-ray continuum spectum.

The deviation from a pure power law component, i.e., the magnitude of the Compton bump, is determined by the interaction between the disk's outer layers and the disk-incident coronal X-ray flux. It is a complex calculation handled by the iterations between \textsc{ptransx} and \textsc{pandurata}. Broadly speaking, the elevated photosphere surfaces at higher $\dot{m}$ result in more coronal photon trajectories intercepted by the disk, and therefore a greater portion of the photons which reach infinity having experienced some Compton downscattering when reflecting off the cooler disk surface.

\item \emph{The power law slope varies non-monotonically with viewer inclination angle.}

The overall flux declines with increasing viewer inclination (i.e., toward more edge-on views), with peak hardness occurring for viewers at $i \sim 60^\circ$, largely independent of the other parameter values. This is due to a larger Compton scattering optical depth through the extended corona for viewers so inclined, relative to the face-on or edge-on observers (who are more likely to see only those photons which escape not having scattered even once). The structure of the corona in our 3D GRMHD simulations develops naturally as a consequence of the buoyancy of magnetic fields. Despite the persistence of point source lamppost-style modeling in the analysis of time-integrated spectra, the spatially-extended coronal structure we find enjoys considerable observational support in the form of variability studies \citep{wil16a, cha19a, zog20a, zog21a}. 

\end{enumerate}

\section{Conclusion}

To sum up: the continuum features are more clearly and dramatically affected by varying the black hole spin and accretion rate than are the K$\alpha$ features. Importantly, the red wing of the Fe K$\alpha$ profile---long used as a proxy for the black hole spin in X-ray reflection spectroscopy studies \citep{mil15a}---actually depends more on the accretion rate than it does on the spin. In fact, we conclude from our parameter study that an observer interested in \emph{measuring} values for these underlying black hole properties using an X-ray spectrum would be best off relying on the relative strengths of the thermal to power law components, and the power law's photon index, in order to constrain the accretion rate and the spin, while using the overall magnitude and the relative strength of the blue wing of the Fe K$\alpha$ line profile to infer the Fe abundance and the viewer inclination angle, respectively.

In this paper, we sought to understand and explain how the \emph{physics} of black hole accretion is affected by the properties of the system and how these physical changes translate into spectral properties---thus the discussion has focused on general trends. However, the predictions we make are quantitative and specific. Moving forward, the template spectra shown in this survey will form the core of a simulation- and physics-based spectral modeling library to be used for the analysis of real X-ray telescope data with \textsc{xspec} \citep{arn96a}.

\acknowledgments

BEK was supported by the U.S. Department of Energy Advanced Simulation and Computing Program's Metropolis Fellowship, through the Los Alamos National Laboratory, and used resources provided by the Los Alamos National Laboratory Institutional Computing Program. Los Alamos National Laboratory is operated by Triad National Security, LLC, for the National Nuclear Security Administration of the U.S. Department of Energy (Contract No. 89233218CNA000001). JHK was partially supported by NSF grants PHY-1707826, AST-1715032, and AST-2009260.

\bibliography{references}{}
\bibliographystyle{aasjournal}

\end{document}